\documentclass[twocolumn]{svjour3}          % twocolumn
\smartqed  % flush right qed marks, e.g. at end of proof
\usepackage[sort&compress,numbers]{natbib}

\usepackage{graphicx}
\usepackage{amssymb}
\usepackage{amsmath}
\usepackage{url}

\usepackage{subfigure}
\usepackage{color}
\newcommand{\new}[1]{\textcolor{black}{#1}}

\newcommand{\Ca}{Ca\textsuperscript{2+}}
\newcommand{\K}{K\textsuperscript{+}}
\newcommand{\R}{\mathbb{R}}

\begin{document}

\title{Geometric slow-fast analysis of 
a hybrid pituitary cell model
%model of a pituitary cell
with
%the effects of 
stochastic ion channel 
dynamics
}
%\subtitle{Do you have a subtitle?\\ If so, write it here}

%\titlerunning{Short form of title}        % if too long for running head

\author{Francesco Montefusco \and Morten Gram Pedersen 
}

%\authorrunning{Short form of author list} % if too long for running head

\institute{F. Montefusco \at
              Dipartimento di Scienze Economiche, Giuridiche, Informatiche e Motorie,
Universita' degli Studi di Napoli Parthenope, Nola (NA), 80035, Italy\\
              \email{francesco.montefusco@uniparthenope.it}           %  \\
%             \emph{Present address:} of F. Author  %  if needed
           \and
           M.G. Pedersen \at
              Department of Information Engineering, University of Padova, Padova, 35131, Italy \\
              \email{mortengram.pedersen@unipd.it}
}

\date{Received: date / Accepted: date}
% The correct dates will be entered by the editor

\maketitle

\begin{abstract}
To obtain explicit understanding of the behavior of dynamical systems, geometrical methods and slow-fast analysis have proved to be highly useful. 
Such methods are standard for smooth dynamical systems, and increasingly used for continuous, non-smooth dynamical systems. 
However, they are much less used for random dynamical systems, in particular for hybrid models with discrete, random dynamics.
Indeed, the analysis of such systems has typically been done by studying the corresponding deterministic system and considering how noise perturbs the deterministic geometrical structures.
Here we propose a geometrical method that works directly with the hybrid system. We illustrate our approach through an application to
%Understanding the role of various ion channels provides valuable and testable insights in the resulting different electrophysiological responses of excitable cells (for example continuous spiking or bursting) and the downstream processes, e.g. calcium-regulated exocytosis. Here, we devise and analyze 
a hybrid pituitary cell model in which %by coupling 
the stochastic dynamics of very few active large-conductance potassium (BK) channels is coupled to
%with those 
a deterministic model of the other ion channels and calcium dynamics. 
To employ our geometric approach, we exploit the slow-fast structure of the model.
The random fast subsystem is analyzed by considering discrete phase planes, corresponding to the discrete number of open BK channels, %form the fast subsystem 
and stochastic events %determine the 
correspond to jumps between these planes. 
The evolution within each plane can be understood from nullclines and limit cycles, and
the overall dynamics, e.g., whether the
model produces a spike or a burst, is determined by the location at which
the system jumps from one plane to another.
Our approach is generally applicable to other scenarios to study discrete random dynamical systems defined by hybrid stochastic-deterministic models.
\keywords{Discrete noise \and random dynamical systems \and hybrid modeling \and ion channels \and action potentials \and bursting electrical activity}
% \PACS{PACS code1 \and PACS code2 \and more}
%\subclass{34A38 \and 34F05 \and 37N25 \and 92-10}
\end{abstract}

\section{Introduction}
\label{intro}

Biological systems are often influenced by discrete, stochastic events, and are therefore appropriately described by hybrid stochastic-deterministic models. Whereas smooth dynamical systems are routinely studied with geometrical techniques, this is still not the case for random dynamical systems such as the ones emerging from hybrid models. The main purpose of the present work is to propose a geometrical method for analyzing hybrid systems, which we illustrate with an application to a hybrid model of cellular electrophysiology.

Many cell types rely on electrical activity  to transduce stimuli to signals to be communicated to other cells. In particular, most endocrine cells release hormones as a result of calcium influx via voltage-sensitive \Ca-channels that open during electrical activity, which triggers calcium-dependent exocytosis of hormone-con\-tain\-ing vesicles \cite{barg03,burgoyne03,stojilkovic05a,pedersen11b,pedersen17}. 
Pituitary endocrine cells, such as the prolactin-secreting lactotrophs studied here, are generally small, and hence stochastic ion channel dynamics can have a large influence on the electrical patterns that they exhibit, and hence on the amount of hormone being released \cite{richards20,fazli21}. 

Large-conductance potassium (BK) channels play a particular important role in lactotrophs, since they can promote so-called bursting electrical activity, where small action potentials fire from a depolarized plateau \cite{van-goor01,tabak07}. This effect is biologically relevant, since bursting leads to higher rates of secretion than simpler action potential firing \cite{van-goor01,tagliavini16,stojilkovic10}. Previous mathematical deterministic modeling and slow-fast analyses have provided insight into how BK channels promote such bursting \cite{tabak07,vo10}. 

However, very few BK channels are active, and  mathematical models must therefore take this stochastic and discrete aspect into account. 
\new{The discrete aspect is particular for BK channels due to their scarcity, and indeed analyses of the effects of stochastic 
dynamics of other types of ion channels %dynamics 
have typically assumed that the system could be described by stochastic differential equations \cite{fox94,fox97,kuskebaer,devries00,pedersen05b,pedersen07a,pedersen07,goldwyn11}.}

A recent simulation study \cite{richards20}  provided some insight \new{into the role of discrete stochastic BK dynamics for shaping electrical activity in pituitary cells}. The authors found numerically that 
%\new{depending on the timing, }
stochastic \new{opening of a} BK channel 
\new{increases the probability of observing a burst if the event happens during the action potential (AP) upstroke or near the AP peak, but lower the burst probability if it occurs after the AP peak. The opposite was observed  for stochastic BK channel closing events. }
%dynamics can trigger bursting... \fbox{more}

Here we extend the work by Richards~et~al.~\cite{richards20} by considering a biologically more correct formulation of the control of BK channels \cite{montefusco17}, and we provide a detailed mathematical analysis of how stochastic opening and closing of discrete BK channels may or may not lead to a burst. Our analysis is based on casting the model in a %formulation 
%that has some aspects of a 
random-dynamical-system formulation \cite{arnold98} combined with geometric analysis.
%
%Analysis of discrete stochastic/hybrid dynamical models \cite{richards20,barabash20,fazli21}
%
We note that since the BK channel transition probabilities depend on the membrane potential $V$, the discrete-noise model that we study is not a so called
blinking system where the stochastic dynamics is independent of the deterministic part, i.e., a system purely driven by a discrete random process, e.g., a Markov Chain  \cite{hasler13,barabash20}.

%\fbox{Stochastic hybrid systems (control notation...)}
\new{Our analysis also differs from traditional geometrical analyses of discrete stochastic models of cellular electrical activity, which typically analyze the deterministic model with  tools from the theory of (smooth) dynamical systems, and then consider noise as perturbations that ``pushes" the system around in phase space \cite{richards20,fazli21}.}
%Based on deterministic tools - typically noise pushes the system  \cite{richards20,fazli21}. 
Here we will show that it is important to understand when and where the ``pushes", i.e. the stochastic events, occur. 
\new{This insight can be obtained by, first, taking advantage of the slow-fast structure of the model, and, then, considering a family of}
%. We will consider discrete jumps of 
nullclines \new{lying in discrete phase planes corresponding to the discrete number of open BK channels. 
Together these planes form the fast-subsystem %model 
phase space, and stochastic events correspond to jumps between these planes.}
The %, which determine the 
dynamics of the system \new{within each plane is determined by geometrical structures such as the nullclines until the next stochastic event, and insight into the overall dynamics can be understood by considering where stochastic jumps between planes occur with respect to the nullclines.}

%% The Appendices part is started with the command \appendix;
%% appendix sections are then done as normal sections
%% \appendix

\section{Methods}
\label{methods}
We devise and analyze a hybrid version of the model by Tabak~et~al.~\cite{tabak07},
%
%The deterministic part of the model is 
\begin{align}
%C_m\, 
\dot V &= -\big(I_{Ca}(V)+I_{Kv}(V,n)+I_{SK}(V,\new{Ca_c}) \nonumber \\
&\qquad\qquad\qquad\ +I_{BK}(V,{\new{Ca_{loc}}})+I_{L}(V)\big)/C_m, \label{dV}\\
\dot n &= (n_{\infty}(V)-n)/{\tau_n}, \label{dn}\\
\dot {Ca_c} &= f (\alpha\, I_{Ca}(V) - k_c\,  \new{Ca_c}),  \label{dCa}  
\end{align}
% \begin{eqnarray}
% \dot V &=& \left(I_{Ca}(V)+I_{Kv}(V,n)+I_{SK}(V,c)+I_{BK}(V,{\new{Ca_{loc}}})+I_{leak}(V)\right)/C_m, \\
% \dot n &=& (n_{\infty}(V)-n)/{\tau_n},\\
% \dot {Ca_c} &=& f (\alpha\, I_{Ca} - k_c\,  \new{Ca_c}),
% \end{eqnarray}
where overdots indicate differentiation with respect to time $t$. $V$ is the cellular membrane potential, $n$ is a gating variable for the voltage sensitive \K\  current ($I_{Kv}$),  and
$\new{Ca_c}$ is the free cytosolic \Ca\ concentration. 
$I_{Ca}$ %, $I_{Kv}$ 
and $I_{L}$ are (deterministic) voltage dependent %membrane 
\Ca\ and leak currents, respectively, $I_{SK}$ is the (deterministic) \Ca-gated small-conductance \K\ (SK) current, while $I_{BK}$ represents the stochastic BK current, which is a function not only of $V$, but also of 
%$Ca_{loc}$, 
%\fbox{other notations as $Ca_l$, $Ca_{loc}$, $Ca_{BK}$???}, 
%denoting 
the local \Ca\ concentration, $\new{Ca_{loc}}$, at  each BK channel, which depends on the surrounding open \Ca\ channels (as explained below).
$C_m$ is the membrane capacitance, 
$\alpha$ changes current to flux, $k_c$ is the \Ca\ removal rate and $f$ is \new{the ratio of} %the 
free-to-bound \Ca. %\ concentration. 

The deterministic currents are modeled as
%\begin{eqnarray}
\begin{align}
I_{Ca}(V)&=g_{Ca}\,m_{CaV,\infty}(V)(V-V_{Ca}),\\
I_{Kv}(V,n)&=g_K\,n\,(V-V_K),\\
I_{SK}(V,Ca_c)&=g_{SK}\,s_{\infty}(Ca_c)(V-V_{K}),\\
I_{L}(V)&=g_{L}\,(V-V_{L}),
\end{align}
% I_{Ca}&=&g_{Ca}m_{CaV,\infty}(V)(V-V_{Ca}),\\
% I_K&=&g_Kn(V-V_K),\\
% I_{SK}&=&g_{SK}s_{\infty}(Ca_c)(V-V_{K}),\\
% I_{leak}&=&g_{l}(V-V_{l}),
%\end{eqnarray}
where $g_X$ represents the whole-cell conductance of the channel $X$ and $V_X$ the corresponding reversal potential.
The steady-state %voltage 
activation \new{functions}, $m_{CaV,\infty}$ and $n_{\infty}$, are described by Boltzmann functions,
\begin{align}
m_{CaV,\infty}(V)&=\frac{1}{1+ \exp((v_{m}-V)/s_m)},\\
n_{\infty}(V)&=\frac{1}{1+ \exp((v_{n}-V)/s_n)},    
\end{align}
% m_{CaV,\infty}(V)&=&\frac{1}{1+ \exp((v_{m}-V)/s_m)},\\
% n_{\infty}(V)&=&\frac{1}{1+ \exp((v_{n}-V)/s_n)},
% \end{eqnarray} 
and the steady-state calcium\new{-dependent} activation \new{function}, $s_{\infty}$, by
\begin{equation}
    s_{\infty}(Ca_c)=\frac{Ca_c^{2}}{Ca_c^2+k_s^2}.
\end{equation}
% s_{\infty}(Ca_c)&=&\frac{Ca_c^{2}}{Ca_c^2+k_s^2}.
% \end{eqnarray}

%The differential equation for the cytosolic Ca$^{2+}$ concentration is
%\begin{eqnarray}
%\frac{d[Ca]}{dt}&=&-f_c(\alpha I_{Ca}+k_c[Ca]).
%\end{eqnarray}
The stochastic current $I_{BK}$ is modeled by
\begin{equation}
\begin{split}
I_{BK}&= \sum_{i=1}^{n_{BK}} \new{\bar g_{BK} \mathbf{1}_{O_{BK,i}}} %p_{O_{BK},i}(V,Ca)
(V-V_K) \\
&=\new{\bar g_{BK} m_{BK}(V-V_K)},   
\end{split}
\end{equation}
where $\new{\bar g_{BK}}$ is the  BK \new{single-}channel conductance, $n_{BK}$  %\fbox{check notation} 
the number of BK channels and 
%$p_{O_{BK},i}$ the open probability of the $i$-th BK channel.
\new{$\mathbf{1}_{O_{BK,i}}$ an index function that equals 1 if the $i$-th BK is open and 0 otherwise, so that $m_{BK}$ indicates the total number of open BK channels.} 
In order to compute \new{the state of the BK channels,} %$p_{O_{BK},i}$, 
we exploit %for the BK channel 
a model of single-channel gating with two states (closed and open), whose dynamics are regulated globally via membrane potential $V$, and locally via the \Ca\ nanodomains below the mouth of \new{stochastic} %the 
CaV channels surrounding the single BK \new{channel} % (i.e. $Ca_{loc}$), as described in our previous works~
\cite{montefusco17,montefusco19}, forming an ion channel BK-CaV complex with a stoichiometry of 1-4 CaV channels per BK channel~\cite{berkefeld10,suzuki13}.
%Large conductance BK potassium channels located in complexed with voltage-sensitive \Ca\ (CaV) currents are described as \cite{montefusco17,montefusco19}
%\fbox{...XXX}
Therefore, in order to model the stochastic gating of BK
channels, 
we describe the transition from one state (closed or open) to another for each BK channel and the surrounding CaVs ($C_X^t$ corresponds to the closed state and $O_X^t$ to the open state of the channel $X$ at time $t$). Then, we define the transition matrix for the single BK,
\begin{equation}
\begin{split}
Q_{BK}
&=\begin{bmatrix}
    P\left(C_{BK}^{t+ \Delta t} |C_{BK}^t\right) &\quad  P\left(C_{BK}^{t+ \Delta t} |O_{BK}^t\right) \\ 
    P\left(O_{BK}^{t+ \Delta t} |C_{BK}^t\right) &\quad  P\left(O_{BK}^{t+ \Delta t} |O_{BK}^t\right) 
  \end{bmatrix}\\
&=\begin{bmatrix}\label{Q_BK}
    1-k^+ \Delta t  &\quad k^- \Delta t \\
    k^+ \Delta t&\quad 1-k^- \Delta t \\
 \end{bmatrix},    
\end{split}   
\end{equation}
and for each CaV surrounding the BK channel,
\begin{equation}
\begin{split}
Q_{CaV}
&=
\begin{bmatrix}
    P\left(C_{CaV}^{t+ \Delta t} |C_{CaV}^t\right) &\quad P\left(C_{CaV}^{t+ \Delta t} |O_{CaV}^t\right) \\ 
    P\left(O_{CaV}^{t+ \Delta t} |C_{CaV}^t\right) &\quad P\left(O_{CaV}^{t+ \Delta t} |O_{CaV}^t\right) 
    % P[C_{CaV},t+ \Delta t |C_{CaV},t] & P[C_{CaV},t+ \Delta t |O_{CaV},t] \\ 
    % P[O_{CaV},t+ \Delta t |C_{CaV},t] & P[O_{CaV},t+ \Delta t |O_{CaV},t] \\ 
  \end{bmatrix}\\
&=\begin{bmatrix}\label{Q_CaV}
    1-\alpha \Delta t  &\quad \beta \Delta t \\
    \alpha \Delta t&\quad 1-\beta\Delta t
  \end{bmatrix},   
  \end{split}  
\end{equation}
where the elements  correspond to the transition probabilities between the indicated states %from state $j$ to state $i$ 
in the time interval $[t, t+\Delta t]$, provided that $\Delta t$ is small. Hence, $\alpha$ and $\beta$ represent the voltage-dependent \Ca\ channel opening %rate 
and closing rate\new{s, respectively, and are given by}
%Since we assume instantaneous CaV activation for this type of cell (CaV activation is much faster than the other dynamics, i.e. $\tau_{CaV}=1.25$ ms) $\alpha$ and $\beta$ are given as in 
\cite{sherman90,montefusco17}%,
\begin{align}
\alpha(V) &= \frac{m_{CaV,\infty}(V)}{\tau_{CaV}},\\
\beta(V) &= \frac{1-m_{CaV,\infty}(V)}{\tau_{CaV}}.
\end{align}
\new{In equation \eqref{Q_BK},}
$k^+$ and $k^-$ are the voltage and %calcium
\Ca-dependent opening and closing rates \new{for BK channels}, %respectively, 
and are modeled as in \cite{montefusco17},
\begin{align}
k^+(V,Ca_{loc}) &= w^+(V)f^+(Ca_{loc})\nonumber \\
&= w_0^+e^{-w_{co}V} \frac{1}{1+\left(\frac{K_{co}}{{Ca_{loc}}}\right)^{n_{co}}},\\
k^-(V,Ca_{loc}) &= w^-(V)f^-(Ca_{loc}) \nonumber \\
&= w_0^-e^{-w_{oc}V} \frac{1}{1+\left(\frac{Ca_{loc}}{{K_{oc}}}\right)^{n_{oc}}},
\end{align}
with $Ca_{loc}$ determined by the \new{number of} %state of the 
surrounding open CaV channels, $n_{CaV_{{BK},o}}$, %(maximum 4 CaVs for each BK-CaV complex~\cite{berkefeld10,suzuki13}):
\begin{equation}
Ca_{loc}={n_{CaV_{{BK},o}}} Ca_{o} +Ca_c, 
\end{equation}
with
\begin{equation}
Ca_o = \frac{i_{Ca}}{8 \pi r D_{Ca}F}\exp{\left[\frac{-r} {\sqrt{\frac{D_{Ca}}{k^+_B[B_{total}]}}}\right]}.
\end{equation}
Here,
$i_{Ca} = \bar g_{Ca}(V-V_{Ca})$ is the single-channel \Ca\ current, $r$ the distance between CaVs and \new{the} BK channel in a single BK-CaV complex, $D_{Ca}$ the \Ca\ diffusion constant, $[B_{total}]$ the total amount of \Ca\ buffer, and $k_B^+$ the on-rate of the buffer. %; w
When all the surrounding CaVs are closed (i.e., $n_{CaV_{{BK},o}}=0$), then $Ca_{loc}=Ca_c$. Note that we assume the linear buffer approximation for computing the \Ca\ profile from $n$ \new{open} channels by superimposing $n$ nanodomains found for single, isolated CaVs.

\new{The hybrid system was solved with a fixed time-step procedure implemented in MATLAB/SIMULINK with $\Delta t =0.01$ ms. 
A third-order
Bogacki-Shampine 
scheme was used to solve the ODEs \eqref{dV}-\eqref{dCa}. The stochastic part of the model, computing the state of the BK-CaV complex, was updated as follows.}
At any time point $t$, 
%the devised model allows computing the state of the BK-CaV complex according the following procedure: 
a random number $\xi$ uniformly distributed on the interval $[0, 1]$ was generated for each of the surrounding CaV channels, and a transition was made based upon the subinterval in which $\xi$ fell; for example, if the CaV channel was open ($O^t_{CaV}$) (see the second column of $Q_{CaV}$ defined by \eqref{Q_CaV}), it remained open if $\xi<1-\beta\Delta t$, otherwise a transition to the closed ($C^{t+\Delta t}_{CaV}$) state occured. 
Similarly, a random number $\eta$ uniformly distributed on the interval $[0, 1]$ for the BK channel was generated, and a transition was made based upon the subinterval that $\eta$ belonged to.
%in which $\eta$ fell. 
%
%This procedure is repeated for every time point to model the stochastic gating of \new{the} BK-CaV complex\new{es in parallel with an Euler scheme  to solve the ODEs \eqref{dV}-\eqref{dCa}, and implemented in MATLAB/SIMULINK.} We use \new{a time step of} $\Delta t=0.01$ ms.

Table~\ref{tab:mod_par} reports the parameter values of the model. 
Matlab code and Simulink schemes implemented for the devised hybrid model with different configurations of the single BK-CaV complex (i.e. different stoichiometries) and their number (i.e. $n_{BK}$)  are provided as Supplementary Material (see Section ``Data availability'').
%\fbox{Table to be UPDATED}.

% \fbox{true?? ho impostato come solver - auto - automatic solver selection -} \\
% \fbox{tra quelli standard che matlab fornisce} \\
% \fbox{\new{MGP: Si riesce a capire quale usa? Mi sembra un po' vago dire "ha scelto MATLAB"...}}

\begin{table*}[tb]
\caption{{\bf Parameter values of the pituitary model}.
%\fbox{CHECK VALUES and UNITS}
}
\centering
\begin{tabular}{c|c|c||c|c|c}
Parameter              & Value            & Unit & Parameter              & Value            & Unit \\
\hline
$C$                      & 10               & pF  & 
$\bar g_{ca}$           & 0.002             & nS\\
$g_{Ca}$               & 2                & nS   &
$V_{Ca}$               & 60               & mV   \\ 
$v_m$                  & -20              & mV   &
$s_m$                  & 12               & mV   \\
$g_K$                  & 3              & nS  &
$V_K$                  & -75              & mV \\
$v_n$                  &-5                & mV &
$s_n$                  & 10               & mV\\
$\tau_n$               & 30               & ms &
$\tau_{CaV}$           & 1.25               & ms \\
$g_{SK}$               & 1.2                & nS &
$k_s$                  & 0.4              & $\mu$M \\
$g_{L}$                & 0.2              & nS &
$V_{L}$                & -50              & mV \\
$f_c$                  &0.01              & -  &
$\alpha$                 & 0.0015           & $\mu$M fC$^{-1}$ \\
$k_{c}$                & 0.12             & ms$^{-1}$ &
$\bar g_{BK}$               & 0.1                & nS  \\
$w_0^-$                & 3.32               & ms$^{-1}$ &
$w_0^+$                & 1.11                & ms$^{-1}$ \\
$w_{oc}$                & 0.022              & mV$^{-1}$  &
$w_{co}$                & $-0.036$            & mV$^{-1}$ \\
$K_{oc}$                  & 0.1               & $\mu$M  &
$K_{co}$                  & 16.6              & $\mu$M \\
$n_{oc}$                  & 0.46               & - &
$n_{co}$                  & 2.33              & - \\
$D_{Ca}$               & 0.250               & ${\mu}$m$^2$ ms$^{-1}$ & 
$F$               & 0.096485      & C $\mu$mol$^{-1}$   \\ 
$k_B$                  & 0.5            & ${\mu}$M$^{-1}$ ms$^{-1}$    &
$B_{total}$                  & 30             & ${\mu}$M \\
$r$         & 0.013 -- 0.030        & $\mu$m & &\\
\hline
\hline
\end{tabular}
% \begin{tabular}{c|c|c}
% Parameter              & Value            & Unit \\
% \hline
% $C$                      & 10               & pF   \\
% $g_{Ca}$               & 2                & nS   \\
% $V_{Ca}$               & 60               & mV   \\ 
% $v_m$                  & -20              & mV   \\
% $s_m$                  & 12               & mV   \\
% $g_k$                  & 3              & nS  \\
% $V_k$                  & -75              & mV \\
% $v_n$                  &-5                & mV \\
% $s_n$                  & 10               & mV\\
% $g_{SK}$               & 1.2                & nS \\
% $k_s$                  & 0.4              & $\mu$M\\ 
% $g_{BK}$               & 1                & nS\\
% $g_{l}$                & 0.2              & nS\\
% $V_{l}$                & -50              & mV\\
% $f_c$                  &0.01              & - \\
% $\alpha$                 & 0.0015           & $\mu$M fC$^{-1}$\\
% $k_{c}$                & 0.12             & ms$^{-1}$\\
% \hline
% \hline
% \end{tabular}
\label{tab:mod_par}
\end{table*}   

\section{Results}
\label{results}

\begin{figure*}
\centering
%\begin{center}
\subfigure[1:1  stoichiometry, $r= 13$ nm]
{\includegraphics[width=.39\textwidth]{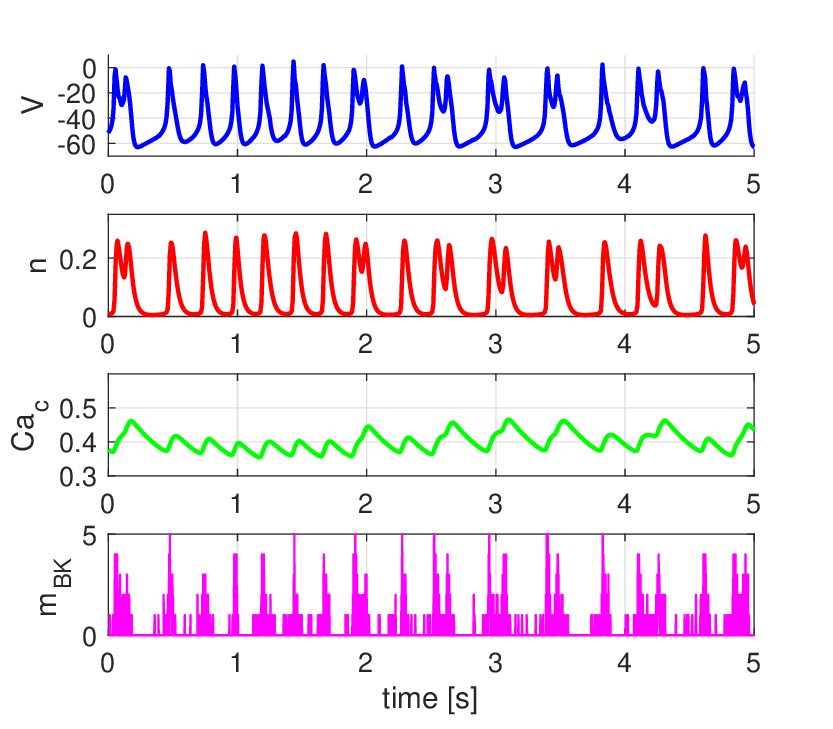}}
\subfigure[1:1  stoichiometry, $r= 30$ nm]
{\includegraphics[width=.39\textwidth]{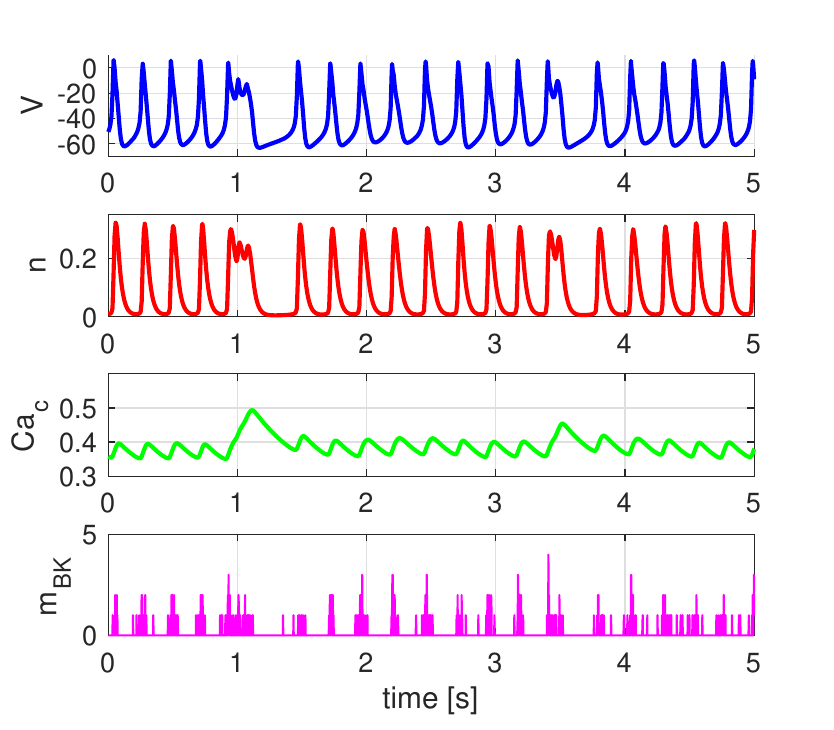}}
\subfigure[1:2 stoichiometry, $r= 13$ nm]
{\includegraphics[width=.39\textwidth]{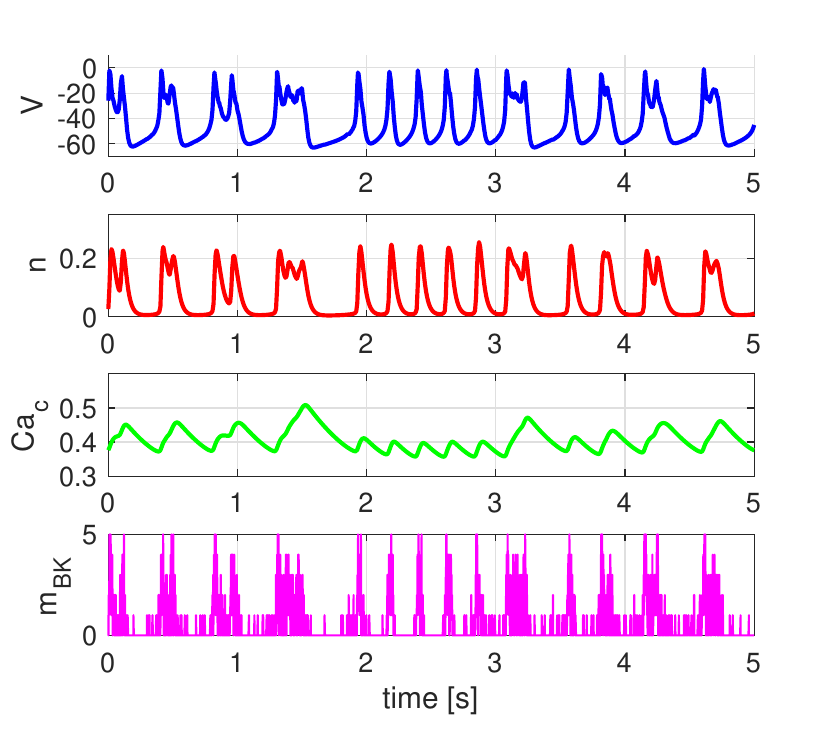}}
\subfigure[1:2  stoichiometry, $r= 30$ nm]
{\includegraphics[width=.39\textwidth]{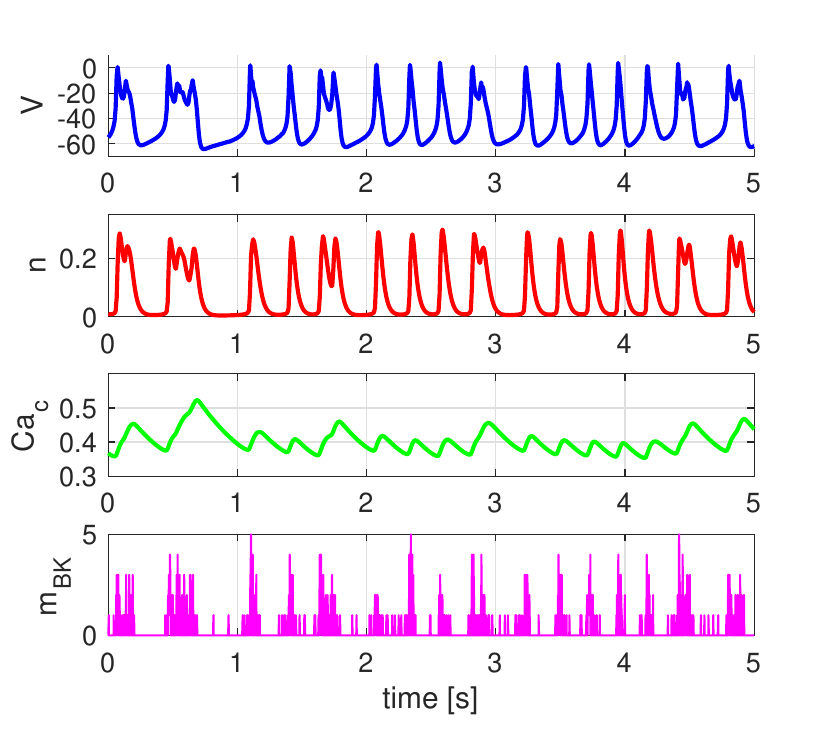}}
\subfigure[1:4  stoichiometry, $r= 13$ nm]
{\includegraphics[width=.39\textwidth]{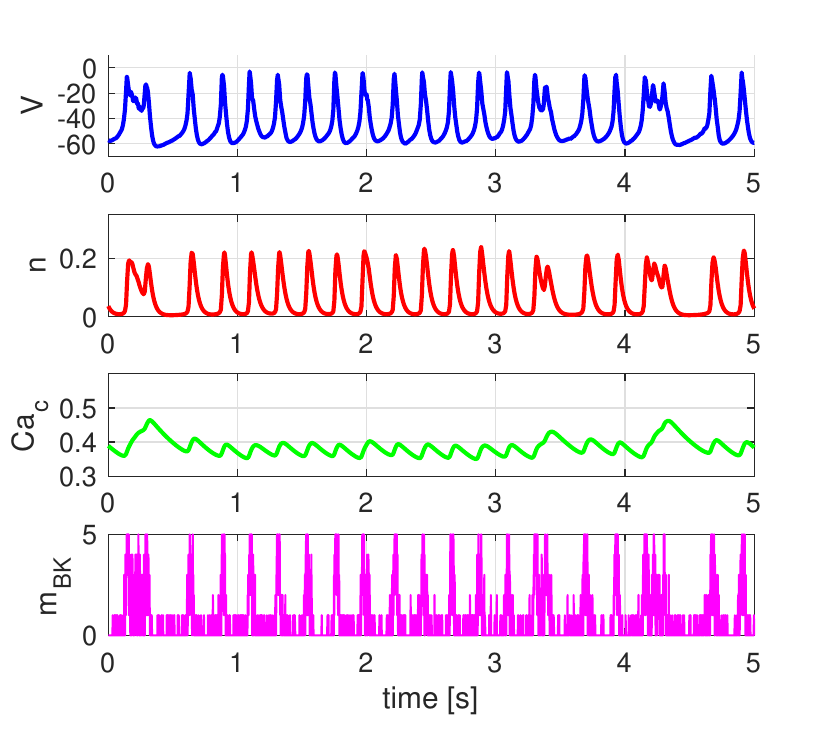}}
\subfigure[1:4  stoichiometry, $r= 30$ nm]
{\includegraphics[width=.39\textwidth]{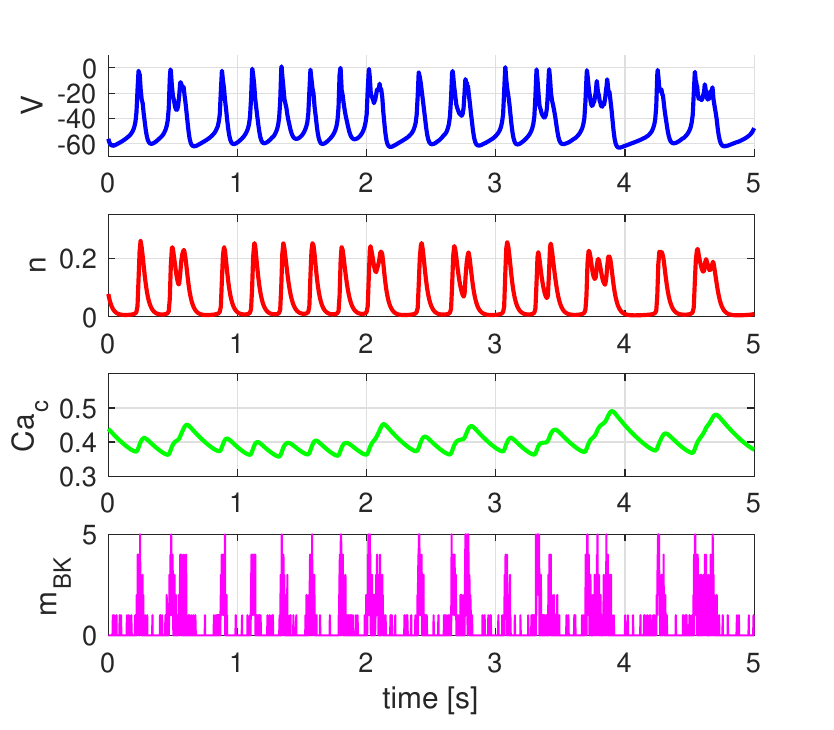}}
%\end{center}
 % \caption[]{\textit{(see next page)}}
 % \end{figure*}
 % \begin{figure} [t!]
    %  \captionsetup{labelformat=adja-page}
    %  \ContinuedFloat
    % \caption%[Figure]
    \caption{The cellular membrane potential $V$ (upper plot in each panel), the gating variable $n$ for the $K_v$ current (second plot), the free cytosolic \Ca\ concentration $Ca_c$ (third plot) and the number of open BK channels (lower plot, $m_{BK}$ open) with respect to time by assuming a number of total BK ($n_{BK}$) equal to 5. For each BKCa-CaV ion channel complex, different  stoichiometries are employed: 1:1 for the first row (panels (a) and (b)); 1:2 for the second row (panels (c) and (d)); 1:4 for the third row (panels (e) and (f)). Also, two different values for the distance $r$ between CaVs and BK channel in a single BK-CaV complex are considered: $r=13$ nm for the first column (panels (a), (c) and (e)); $r=30$ nm for the second column (panels (b), (d) and (f)).} \label{fig:nBK5}
\end{figure*}

The hybrid model produces different kinds of behavior depending on the configuration of the BK-CaV complexes and their number. 
\new{Figure~\ref{fig:nBK5} shows simulated traces for $n_{BK}=5$ BK channels in complexes with 1, 2 or 4 CaVs, which are located either 13 nm or 30 nm from the BK channel of the complex.
The membrane voltage $V$ exhibits both single action potential firing as well as so called bursts where small-amplitude action potentials and voltage fluctuations appear from a depolarized plateau. 
Interesting, bursting appears to be less frequent in the BK-CaV configurations with either many CaVs located close to the BK channel (1:4 stochiometry, $r=13$ nm, Fig.~\ref{fig:nBK5}e), or with few  CaVs at a greater distance from the BK channel (1:1 stochiometry, $r=30$ nm, Fig.~\ref{fig:nBK5}b). 
These two cases correspond respectively to the configurations where we would expect BK channels to open readily or less frequently, as confirmed by the lower traces showing the number of open BK channels.}

%\begin{figure}[htbp]
%\begin{center}
%\includegraphics[width=0.32\textwidth]{Figs/nBK_5_nCaV_1_rBK_13_a.pdf}
%\includegraphics[width=0.32\textwidth]{Figs/nBK_5_nCaV_1_rBK_13_a.pdf}
%\includegraphics[width=0.32\textwidth]{Figs/nBK_5_nCaV_1_rBK_13_a.pdf}
%\includegraphics[width=0.99\textwidth]{Figs/example_burst1_pp}
%\caption{}
%\label{fig:n_BK_5}
%\end{center}
%\end{figure}

% \begin{figure}
% \begin{center}
% \subfigure[1:1  st.; $r= 13$ nm]
% {\includegraphics[width=.327\columnwidth]{Figs/nBK_10_nCaV_1_rBK_13.eps}}
% \subfigure[1:1  st.; $r= 30$ nm]
% {\includegraphics[width=.327\columnwidth]{Figs/nBK_10_nCaV_1_rBK_30.eps}}
% \subfigure[1:1  st.; $r= 60$ nm]
% {\includegraphics[width=.327\columnwidth]{Figs/nBK_10_nCaV_1_rBK_60.eps}}
% \subfigure[1:2  st.; $r= 13$ nm]
% {\includegraphics[width=.327\columnwidth]{Figs/nBK_10_nCaV_2_rBK_13.eps}}
% \subfigure[1:2  st.; $r= 30$ nm]
% {\includegraphics[width=.327\columnwidth]{Figs/nBK_10_nCaV_2_rBK_30.eps}}
% \subfigure[1:2  st.; $r= 60$ nm]
% {\includegraphics[width=.327\columnwidth]{Figs/nBK_10_nCaV_2_rBK_60.eps}}
% \subfigure[1:4  st.; $r= 13$ nm]
% {\includegraphics[width=.327\columnwidth]{Figs/nBK_10_nCaV_4_rBK_13.eps}}
% \subfigure[1:4  st.; $r= 30$ nm]
% {\includegraphics[width=.327\columnwidth]{Figs/nBK_10_nCaV_4_rBK_30.eps}}
% \subfigure[1:4  st.; $r= 60$ nm]
% {\includegraphics[width=.327\columnwidth]{Figs/nBK_10_nCaV_4_rBK_60.eps}}
% \end{center}
% \caption{{$V$, $n$, $c$ and \# BK open with respect to time for $n_{BK}=10$. Panel details are as in Fig.~\ref{fig:nBK5}. }} \label{fig:nBK10}
% \end{figure}

\begin{figure*}
\centering
%\begin{center}
\subfigure[1:1  stoichiometry, $r= 13$ nm]
{\includegraphics[width=.39\textwidth]{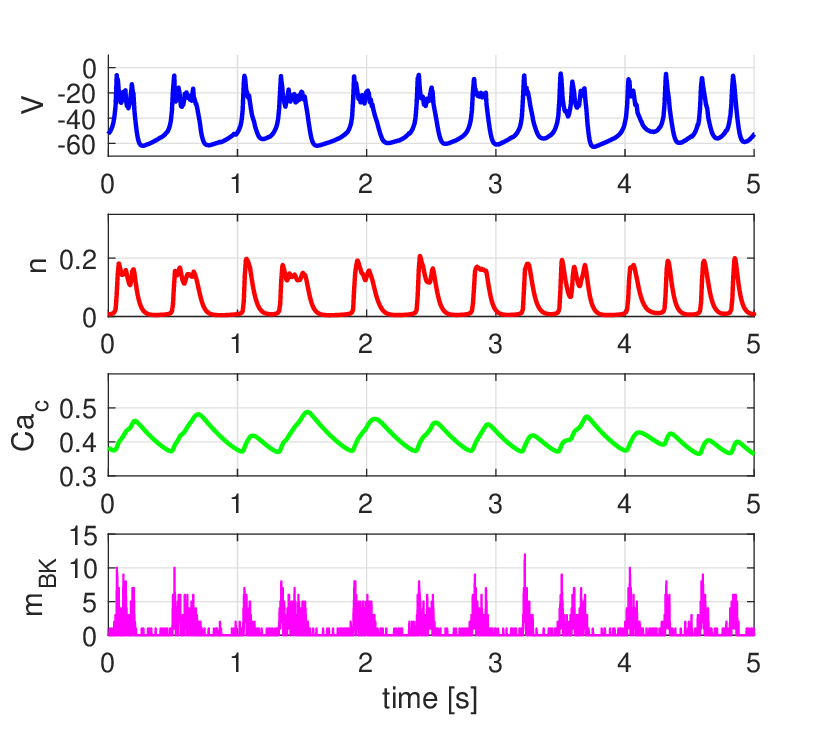}}
\subfigure[1:1  stoichiometry, $r= 30$ nm]
{\includegraphics[width=.39\textwidth]{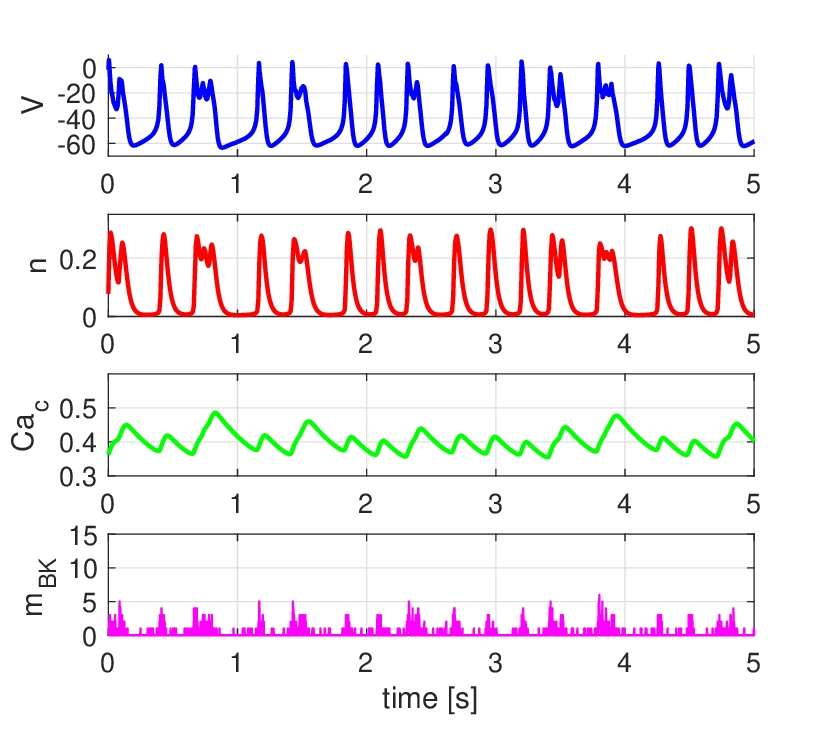}}
\subfigure[1:2 stoichiometry, $r= 13$ nm]
{\includegraphics[width=.39\textwidth]{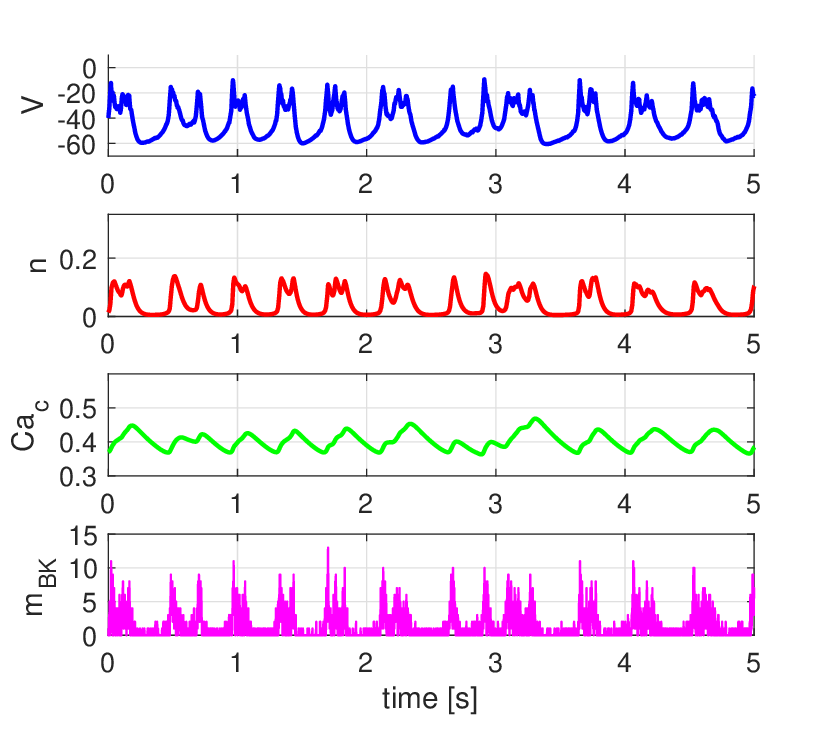}}
\subfigure[1:2  stoichiometry, $r= 30$ nm]
{\includegraphics[width=.39\textwidth]{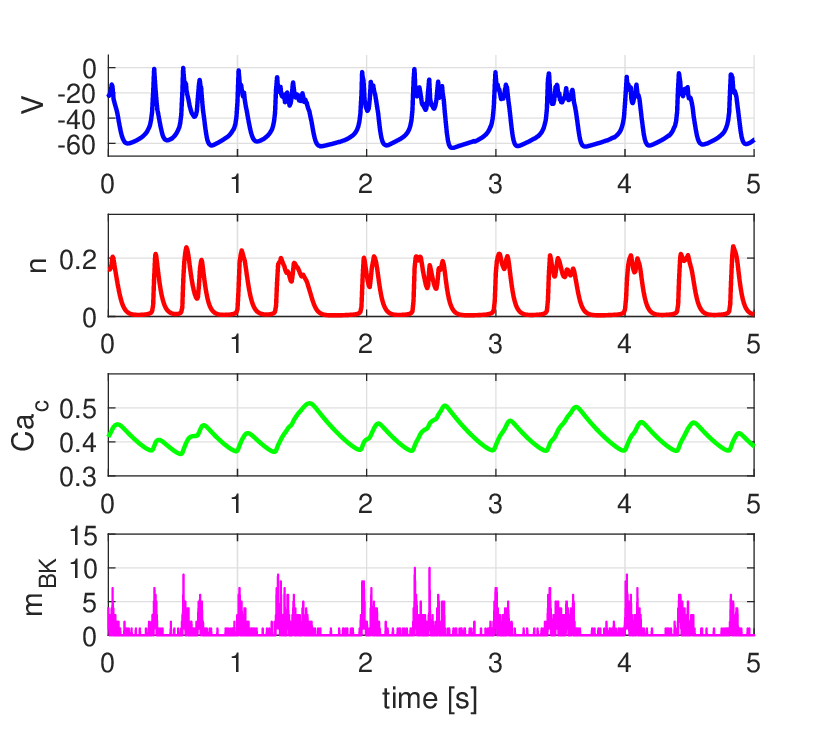}}
\subfigure[1:4  stoichiometry, $r= 13$ nm]
{\includegraphics[width=.39\textwidth]{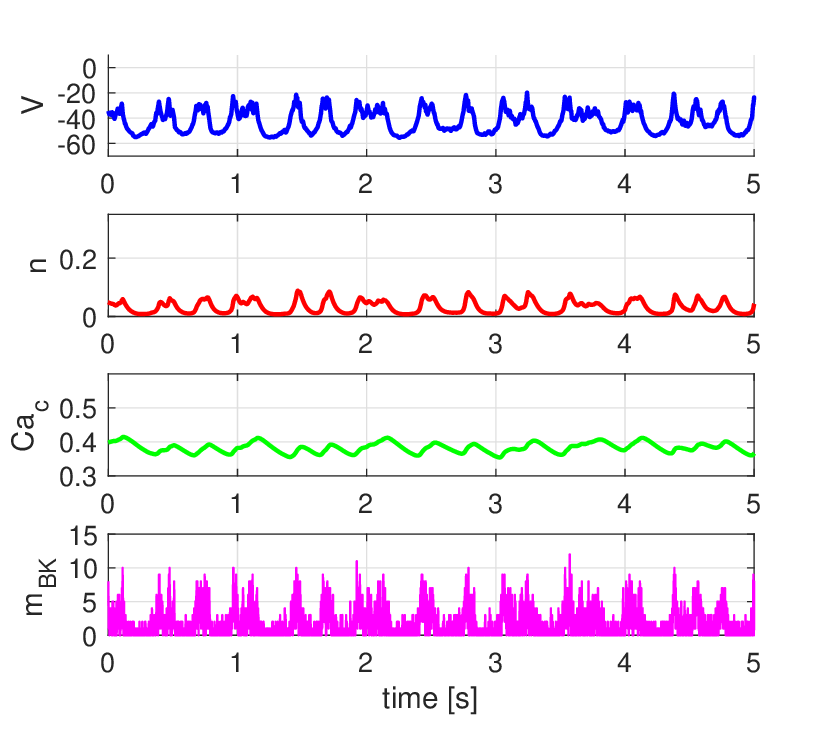}}
\subfigure[1:4  stoichiometry, $r= 30$ nm]
{\includegraphics[width=.39\textwidth]{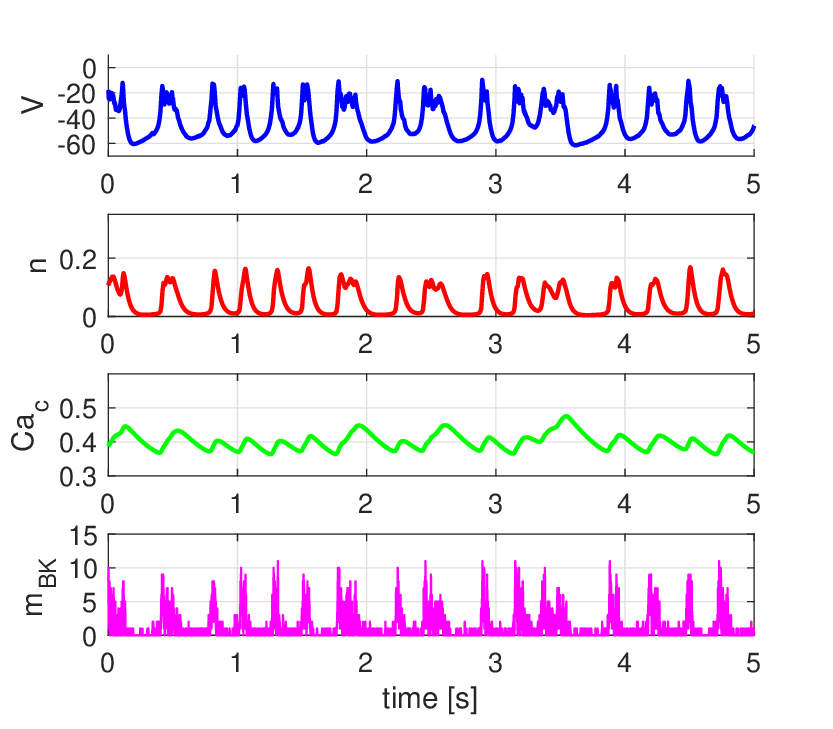}}
%\end{center}
\caption{{%$V$, $n$, 
\new{$V,n,Ca_c$} and $m_{BK}$ %open with respect to 
\new{as functions of} time for $n_{BK}=15$. % channels. %Panel d
Details %are 
as in Fig.~\ref{fig:nBK5}. }} 
\label{fig:nBK15}
\end{figure*}

\new{With more BK channels, bursting seems to be more frequent (Fig.~\ref{fig:nBK15}, $n_{BK}=15$). 
However, as seen most clearly for the configuration with 1:4 stoichiometry and $r=13$ nm (Fig.~\ref{fig:nBK15}e), the larger number of BK channels tend to make the interburst interval more depolarized ($\sim -50$ mV) and the active phase of the burst more hyperpolarized ($\sim -30$ mV), compared to the behavior seen for $n_{BK}=5$ (Fig.~\ref{fig:nBK5}).
}

\new{To understand the differences seen in the stochastic simulations for the various configurations, we first}
aim at \new{obtaining a geometric understanding of} why the model sometimes produce a burst and sometimes produce a spike. \new{We will then take advantage of this insight to explain the behavior seen in the different cases in Figs.~\ref{fig:nBK5} and \ref{fig:nBK15}.}

The time-scale of the gating variable $n$ of the Kv-current is $\tau_n=30$ ms, $m_{BK}$ has time scale of a few ms \cite{cox14,montefusco17}, whereas $Ca_c$ has time scale $1/(f k_c) = 833$ ms, which allow us to treat the system as a slow-fast system with $Ca_c$ as a slow variable, and the remaining variables $(V,n,m_{BK})$ as a hybrid (random) fast subsystem of the full model.

The phase space of this subsystem is composed of $1+n_{BK}$ discrete planes, more specifically -- since $n\in[0,1]$ -- it equals  $\R\times [0,1]\times\{0,\ldots,n_{BK}\}$.
It is useful to consider the $V$ and $n$ nullclines for fixed $m_{BK}$, the number of open BK channels. The $n$ nullcline is independent of $m_{BK}$ and given by $n=n_{\infty}(V)$. The $V$ nullcline, in contrast, depends on $m_{BK}$ and on the (fixed) value of $Ca_c$. 
%As $n_{BK}$ increases, the $V$ nullcline moves downwards in the $(V,n)$ plane.

\begin{figure*}[htbp]
\centering
%\begin{center}
\subfigure[]{\includegraphics[width=0.35\textwidth]{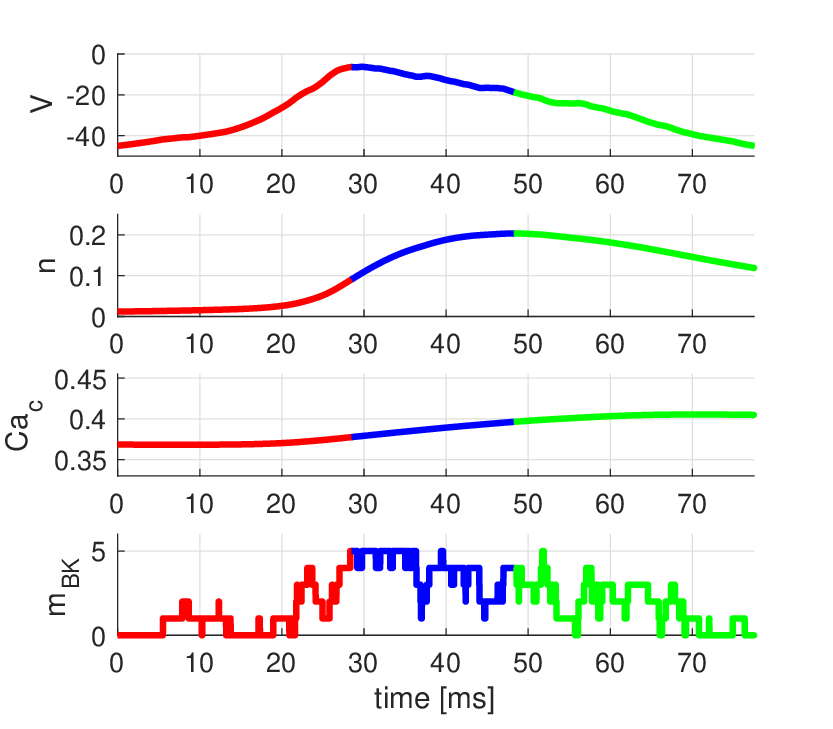}}
\subfigure[]{\includegraphics[width=0.35\textwidth]{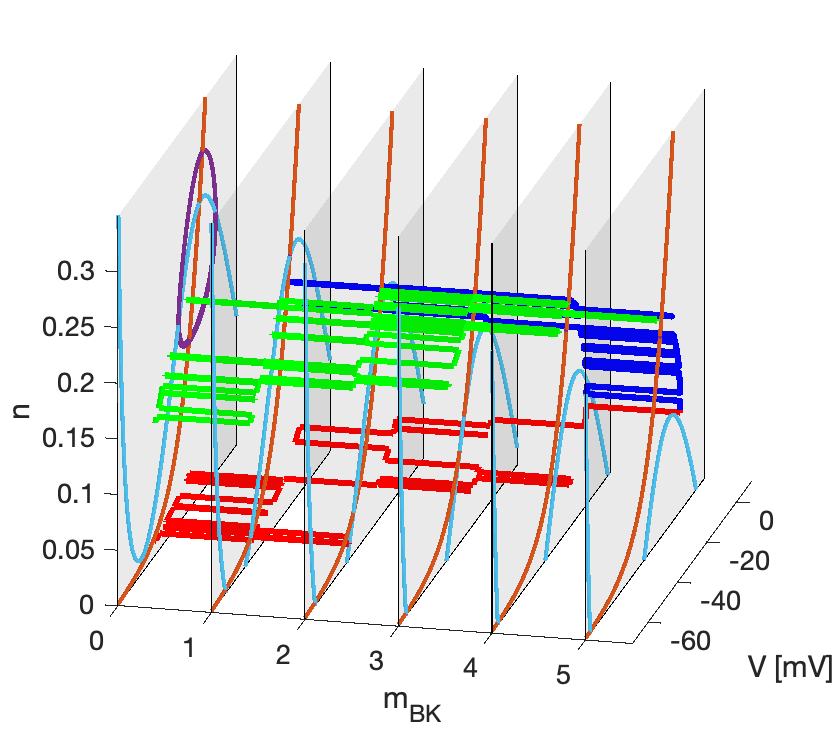}}
\subfigure[]{\includegraphics[width=0.7\textwidth]{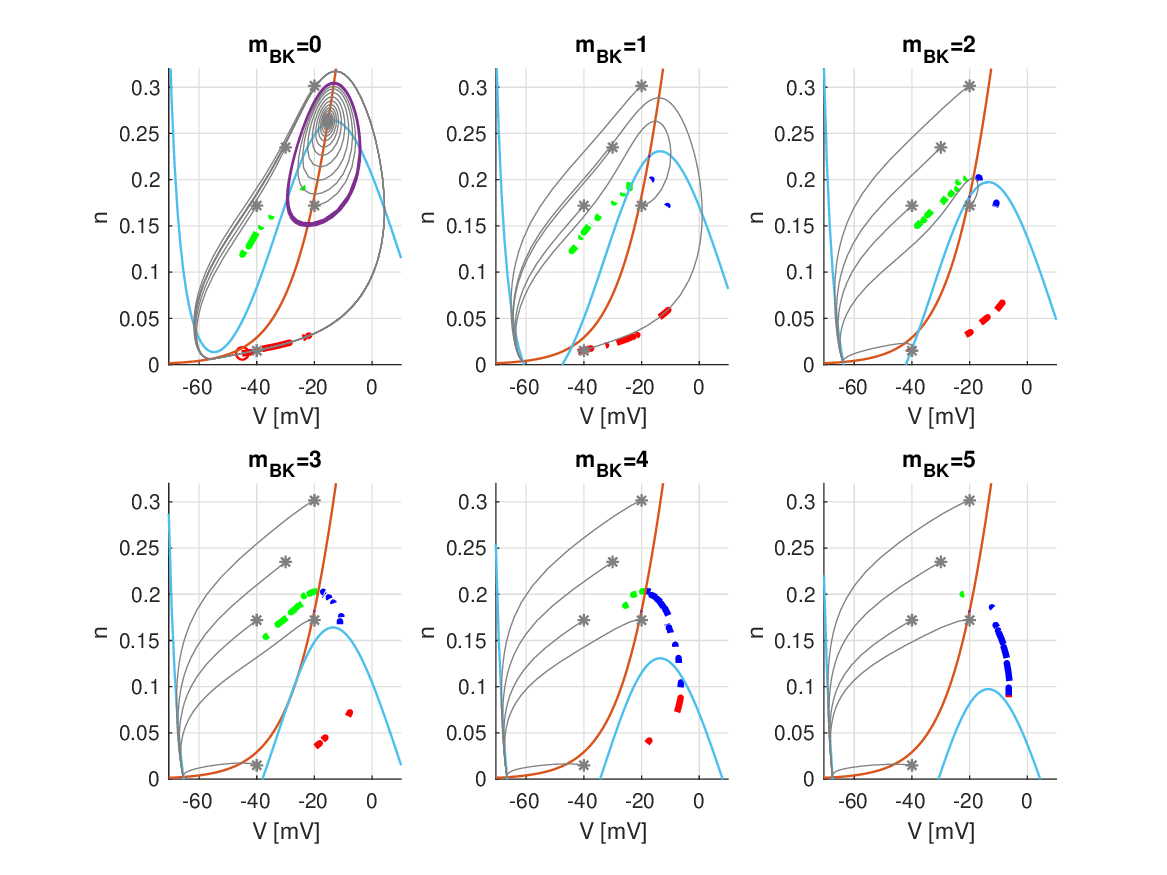}}
%\includegraphics[width=0.89\textwidth]{Figs/example_spike1_pp}
%\end{center}
 % \caption[]{\textit{(see next page)}}
 % \end{figure*}
 % \begin{figure} [t!]
 %     \captionsetup{labelformat=adja-page}
 %     \ContinuedFloat
\caption{{\textbf{\new{Single action potential example.}} %An example of a 
A single action potential (spike) for $n_{BK}=$ 5 BK channels in complexes with 4 CaVs (i.e., 1:4 stoichiometry) located 13 nm ($=r$) from the BK of the complex. %}} 
\textbf{(a)} $V$, $n$, $Ca_c$ and $m_{BK}$ as functions of time. It corresponds to the spike of Fig.~\ref{fig:nBK5}e for $4.64 \le t \le 4.72$. The colors of the curves %are provides to 
indicate different phases of the action potential for easier comparison to  panels (b) and (c).
\textbf{(b)} Projection of the simulation in panel (a) onto the phase space of the fast $(V,n,m_{BK})$ subsystem (colors as in panel (a)). 
The phase space is composed of the gray planes given by constant $m_{BK}$, and random opening and closing of BK channels correspond to jumps between these planes. 
For fixed $m_{BK}$, the $V$ (light blue) and $n$ (red) nullclines are shown for $Ca_c=0.4\,\mu$M.
\textbf{(c)} Phase planes for fixed $m_{BK}$ as indicated and $Ca_c=0.4\,\mu$M, corresponding to the gray planes in panel (b), with $V$ (light blue) and $n$ (red) nullclines. The red circle for $m_{BK}=0$ represent the first sample of the time simulation.
The violet oval for $m_{BK}=0$ indicates the unstable limit cycle. 
Gray asterisks indicate initial conditions for different deterministic orbits obtained by keeping $m_{BK}$ fixed and equal to the value of the corresponding panel. For $m_{BK}=0$ the trajectories starting outside the unstable limit cycle converge to a stable period orbit (gray).
The simulation in panel (a) is projected onto the plane with the corresponding value of $m_{BK}$ using dots with colors as in panel (a).
Note how the system returns to the $m_{BK}=0$ plane to the left of the $V$ nullcline (green points), which forces $V$ to decrease further, ending the action potential. 
%A new spike or burst (as in Figs.~\ref{fig:burst}-\ref{fig:burst_c}) can then develop.
}}
\label{fig:spike}
\end{figure*}

\begin{figure*}[htbp]
\centering
%\begin{center}
\subfigure[]{\includegraphics[width=0.35\textwidth]{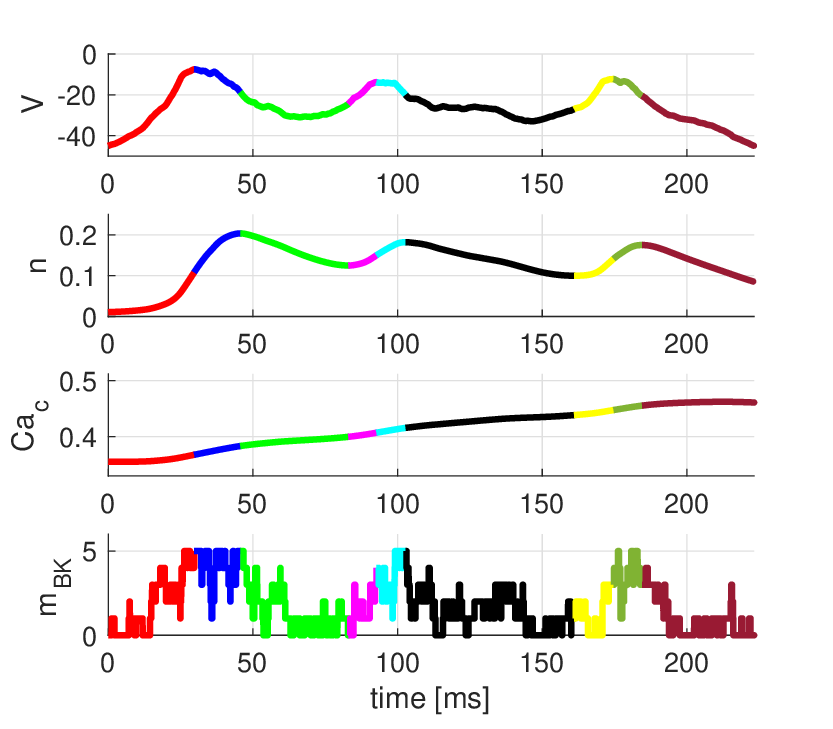}}
\subfigure[]{\includegraphics[width=0.35\textwidth]{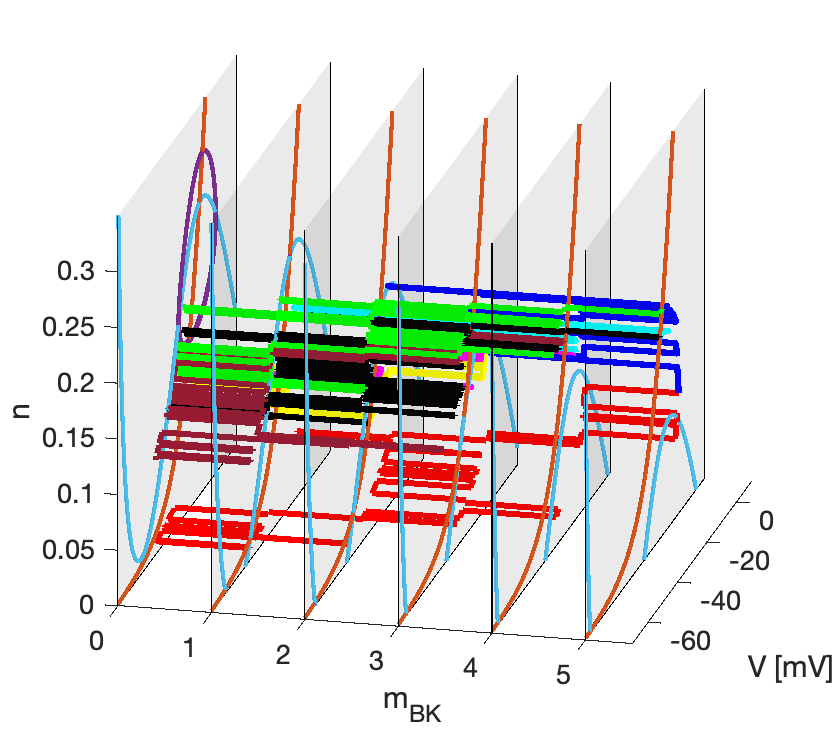}}
\subfigure[]{\includegraphics[width=0.7\textwidth]{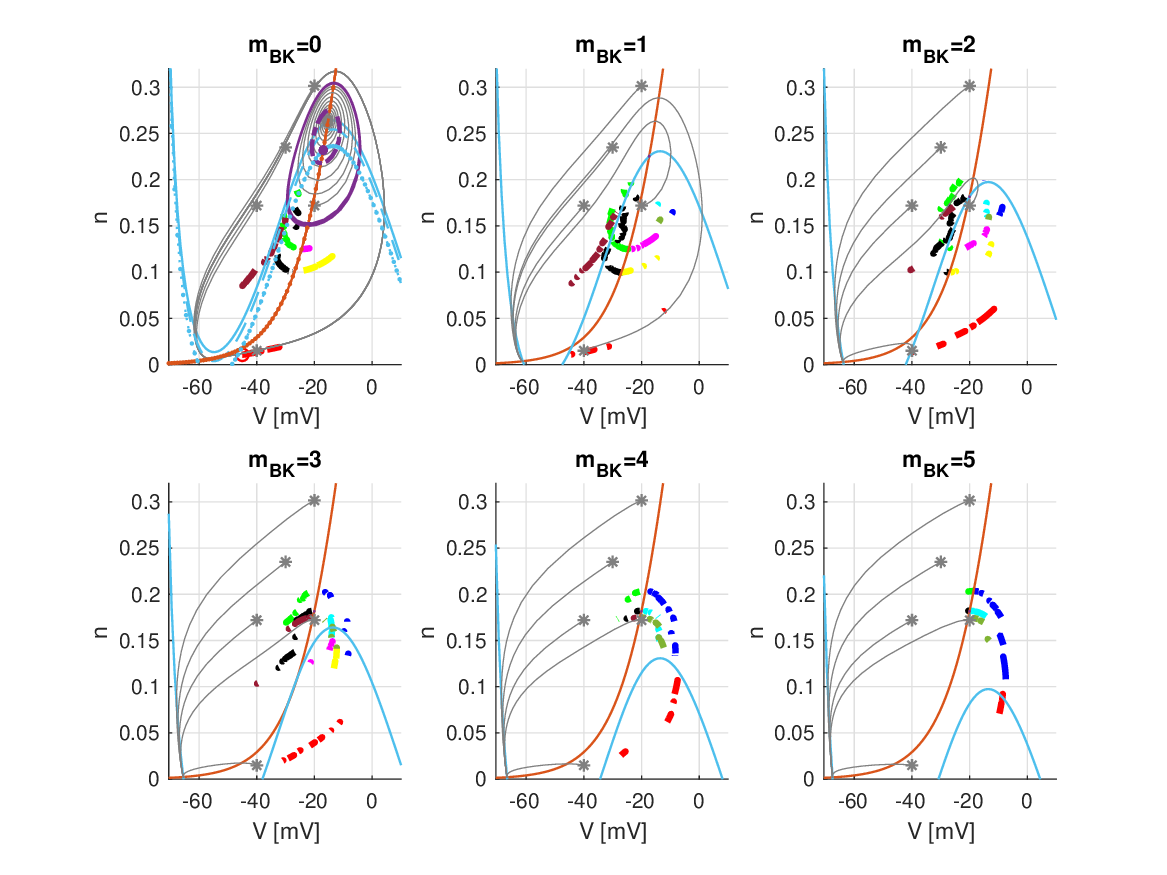}}
%\includegraphics[width=0.89\textwidth]{Figs/example_spike1_pp}
 % \caption[]{\textit{(see next page)}}
 % \end{figure*}
 % \begin{figure} [t!]
 %     \captionsetup{labelformat=adja-page}
 %     \ContinuedFloat
     \caption{{\textbf{An example of a burst %with three peaks 
     for $n_{BK}=5$ %5 BK channels, 
     with 1:4 stoichiometry and $r=13$ nm}}. It corresponds to the burst of Fig.~\ref{fig:nBK5}e for $4.12 \le t \le 4.35$. Legends as in Fig.~\ref{fig:spike}. In the first subplot of panel (c), for $m_{BK}=0$, the $V$-nullcline (light-blue) and the unstable limit cycle (violet) are calculated for $Ca_c=0.4\,\mu$M (solid), $Ca_c=0.42\,\mu$M (dashed) and $Ca_c=0.46\,\mu$M (dotted; the limit cycle is very small with center (-17 mV, 0.23)).
     Note how the system returns to the $m_{BK}=0$ plane  to the right of the $V$-nullclines the first two times (first time, green points to the right of the solid light-blue line; second time, black points to the right  of dashed light-blue line), leading to a new increase of $V$. Only the third time, when $Ca_c$ has increased, moving  the $V$ nullcline further downwards (dotted light-blue line), does the system (brown points) return to the $m_{BK}=0$ plane to the left of the $V$ nullcline, which causes a further decrease of $V$, terminating the burst. 
     %Note how the oval unstable limit cycle for $m_{BK}=0$ reduces by increasing $Ca_c$, approaching to the infinitesimal cycle with center (-17 mV, 0.23) for $Ca_c=0.46\,\mu$M.
     }
\label{fig:burst}
%\end{center}
\end{figure*}

Drawing these nullclines for $n_{BK}=5$ 
%\fbox{
and $Ca_c=0.4\,\mu$M 
%??} 
in each of the planes $\R\times [0,1]\times\{m_{BK}\}$ shows that for $m_{BK}=0$ only a single stable equilibrium exists at $V\approx -15$ mV, which is surrounded by an unstable limit cycle and a larger stable limit cycle (Figs.~\ref{fig:spike}bc and~\ref{fig:burst}bc).
Increasing $m_{BK}$, the $V$ nullcline moves downwards and the unstable limit cycle disappears so that the upper stable equilibrium becomes unstable for ${m_{BK}=1}$ and $m_{BK}=2$, where three equilibriums are present: the lower one is stable and the one in the middle is a saddle point. For $m_{BK}\geq 3$ only the lower stable equilibrium is present.

At the beginning of an action potential when the cell is hyperpolarized, the BK channels tend to close so $m_{BK}=0$ most of the time. For this value of $m_{BK}$ the system is attracted to the large limit cycle. 
However, as $V$ increases the opening rate of the channels increases compared to their closing rate and the fast subsystem jumps to the planes with $m_{BK}>0$, eventually reaching the plane with $m_{BK}=5$ \new{(Fig.~\ref{fig:spike}abc)}. 
In parallel to the increase in $m_{BK}$, the Kv gating variable $n$ increases as well, with the result that the fast subsystem is above the $V$ nullcline in the $m_{BK}=5$ plane when the trajectory reaches this plane. 
Consequently, $V$ starts to decrease whereas $n$ still increases. As the cell hyperpolarizes, the BK channels begin to close, eventually followed by a decrease in the $n$ variable.

What distinguishes a spike from a burst is the point at which the trajectory reaches the $m_{BK}=0$ plane. 
If it happens \new{above and to the left of the middle part of} $V$ nullcline, $V$ will continue to decline, terminating the action potential so that a single spike occurred (Fig.~\ref{fig:spike}). 
\new{This is true even if a BK channel should open as this would move the $V$ nullcline downwards and the trajectory would remain above.}

If, on the other hand, the trajectory reaches the ${m_{BK}=0}$ plane below/to the right of the $V$ nullcline \new{as in Fig.~\ref{fig:burst}c}, $V$ starts to increase leading to a second action potential.
This scenario might repeat itself several times with the result that a burst is formed.
During the burst, the \Ca\ level $Ca_c$ increases (Fig.~\ref{fig:burst}a), which moves the $V$ nullcline downwards \new{(Fig.~\ref{fig:burst}c)}. %, %which favors the possibility 
This shift increases the probability that the system hits above/to the left of the $V$ nullcline in the $m_{BK}=0$ plane, which would terminate the burst. That is, slow feedback from \Ca\ contributes to controlling the end of the burst.

\begin{figure}[tbp]
\begin{center}
\includegraphics[width=\columnwidth]{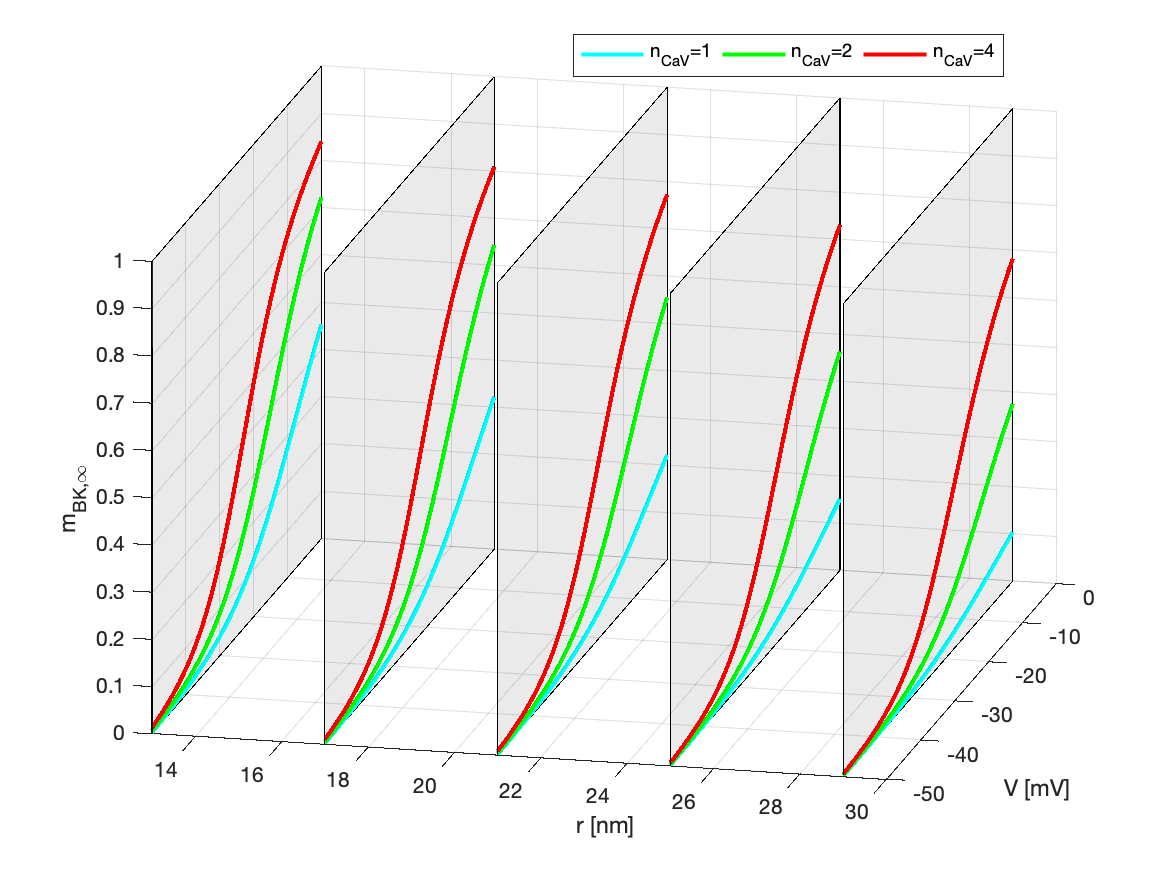}
\caption{Steady-state BK activation function, $m_{BK,\infty}$
%, as modelled in~\cite{montefusco17} 
(see Eq.~29 in~\cite{montefusco17}).}
%\fbox{Is this a Supplementary Figure??}}
\label{fig:minf}
\end{center}
\end{figure}

\new{With less CaVs in the BK-CaVs complexes, or with a greater distance between the CaVs and the BK channel, the  steady-state average fraction of open BK channels decreases for any given $V$ 
%(Fig.~S\ref{fig:spike_b} \fbox{SI Fig??})
(Fig.~\ref{fig:minf}), as expected from the biophysical fact that BK channels are \Ca\ activated, and hence, if a lower \Ca\ concentration is present at the BK channel because of fewer or more distant CaVs in its complex, the open probability decreases.}

\begin{figure*}[htbp]
\centering
%\begin{center}
\subfigure[]{\includegraphics[width=0.35\textwidth]{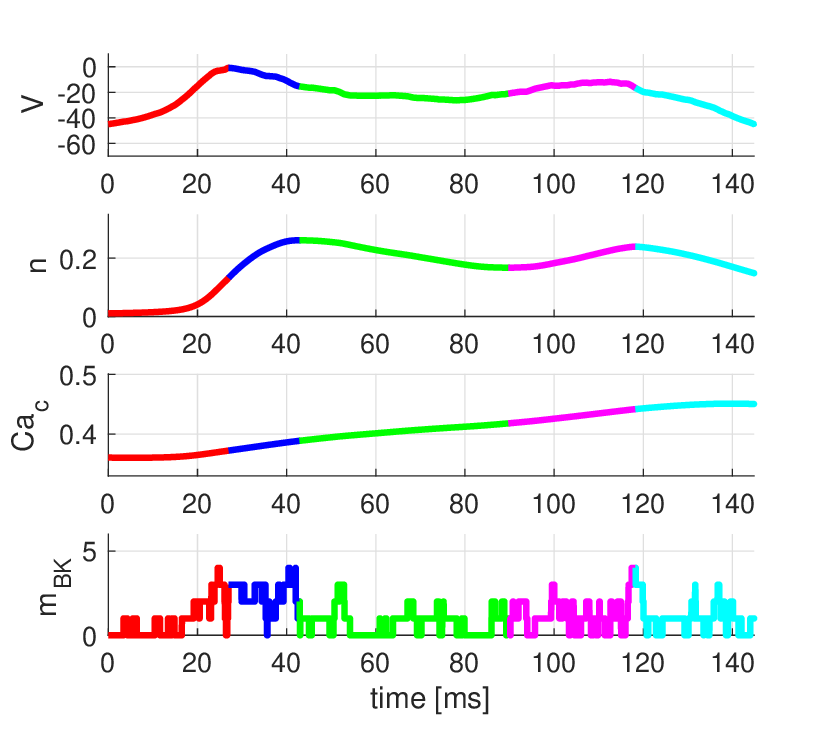}}
\subfigure[]{\includegraphics[width=0.35\textwidth]{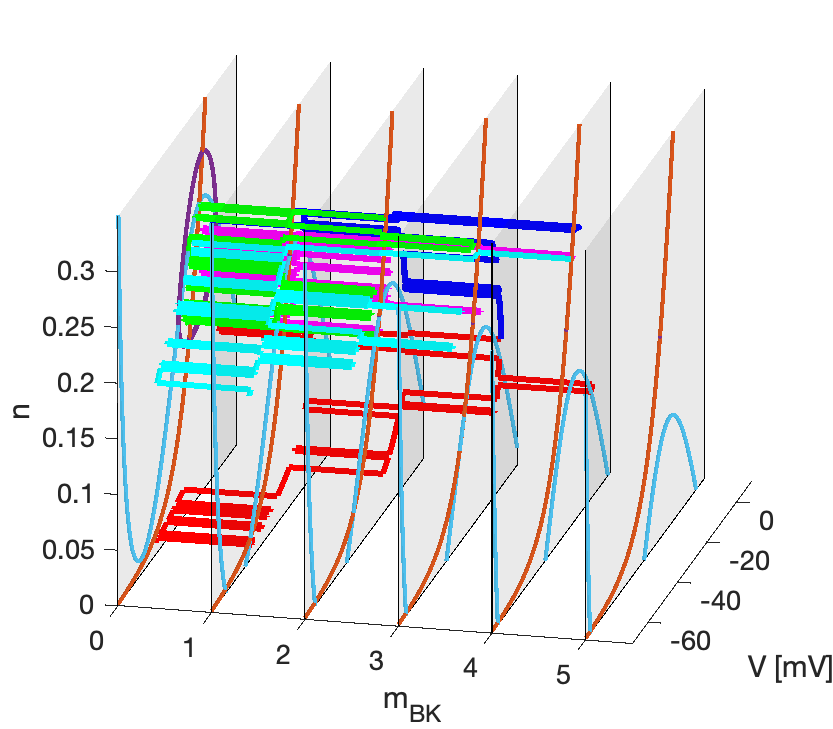}}
\subfigure[]{\includegraphics[width=0.7\textwidth]{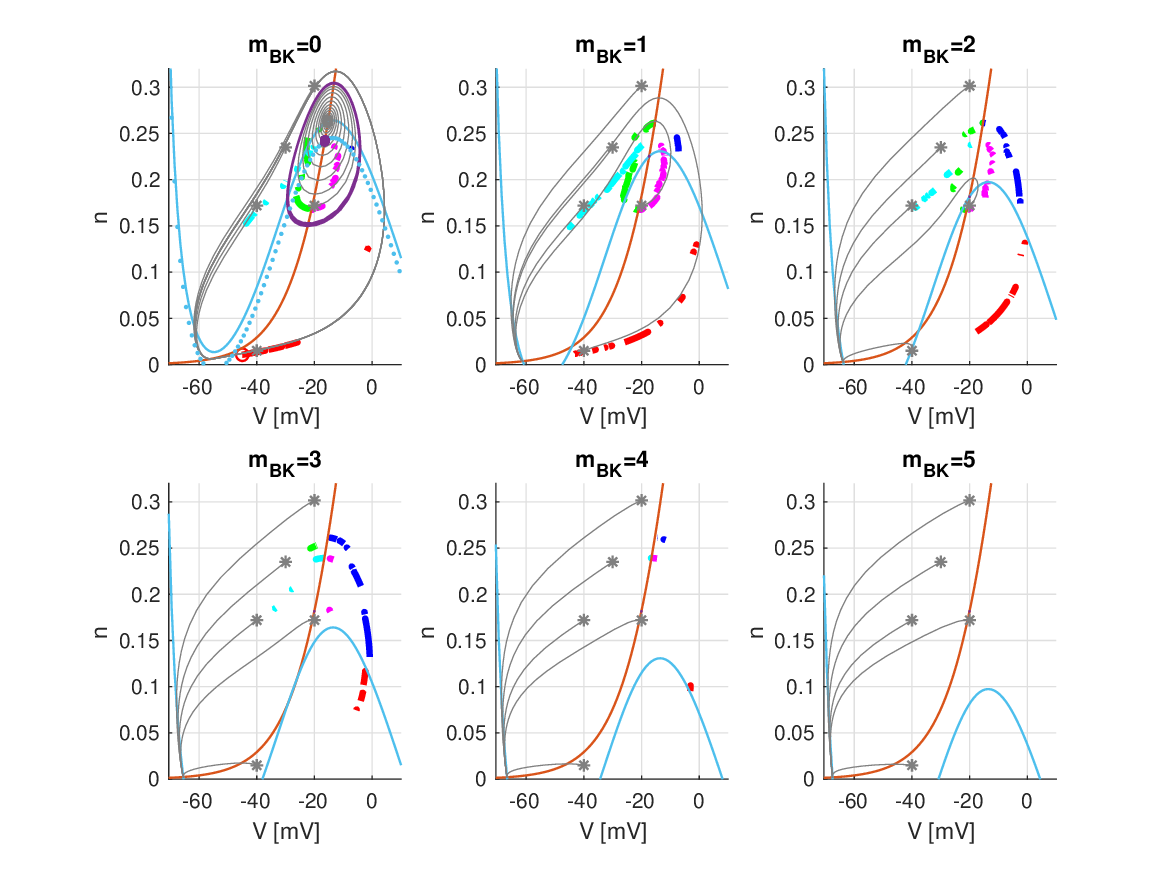}}
 % \caption[]{\textit{(see next page)}}
 % \end{figure*}
 % \begin{figure} [t!]
 %     \captionsetup{labelformat=adja-page}
 %     \ContinuedFloat
     \caption{{\textbf{An example of a burst for $n_{BK}=5$ %channels, 
     with 1:1 stoichiometry and $r=13$ nm}}. It corresponds to the burst of Fig.~\ref{fig:nBK5}a for $4.82 \le t \le 4.96$. Legends as in Fig.~\ref{fig:spike}. 
     Note how the system returns to the $m_{BK}=0$ plane (first subplot of panel (c)) within the unstable limit cycle (green points inside the violet oval) %to the right of the solid blue light $V$-nullcline for $Ca_c=0.4\,\mu$M and within the unstable limit cicle,
     leading to a new increase of $V$ (magenta points). Only the second time, when $Ca_c$ has increased moving  the $V$ nullcline slightly downwards, does the system return to the $m_{BK}=0$ plane to the left of the $V$ nullcline (cyan points to the left of the dotted light-blue line with $Ca_c=0.44\,\mu$M), which causes a further decrease of $V$, terminating the burst. For $Ca_c=0.44\,\mu$M, the unstable limit cycle is also reduced.}
\label{fig:burst_b}
%\end{center}
\end{figure*}

\begin{figure*}[htbp]
\centering
%\begin{center}
\subfigure[]{\includegraphics[width=0.35\textwidth]{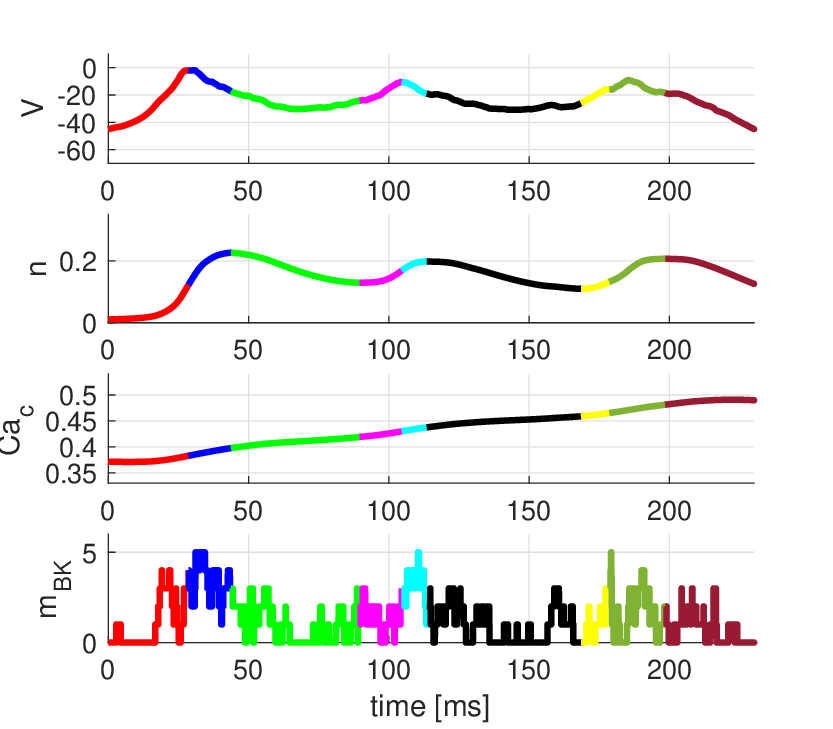}}
\subfigure[]{\includegraphics[width=0.35\textwidth]{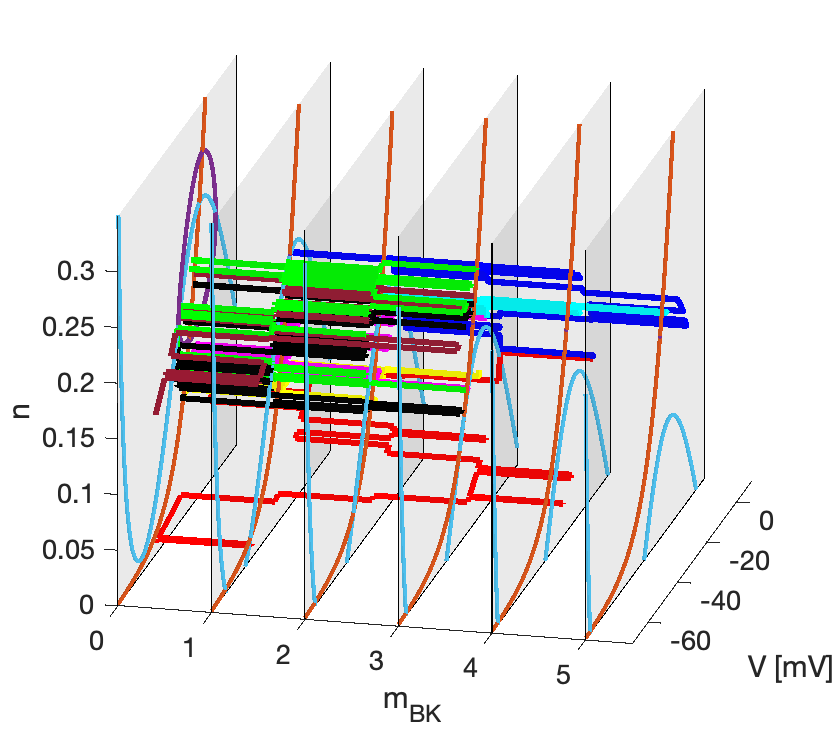}}
\subfigure[]{\includegraphics[width=0.7\textwidth]{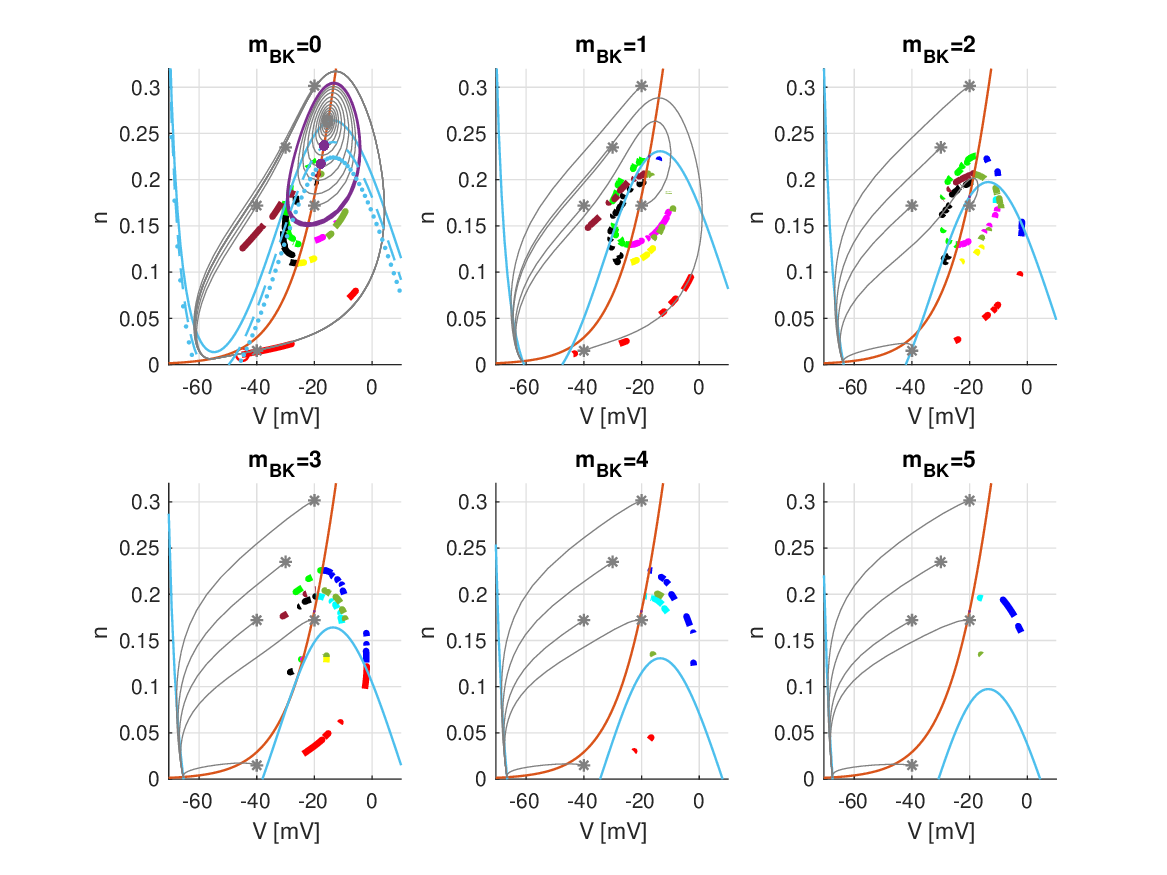}}
%\includegraphics[width=0.89\textwidth]{Figs/example_spike1_pp}
 % \caption[]{\textit{(see next page)}}
 % \end{figure}
 % \begin{figure} [t!]
 %     \captionsetup{labelformat=adja-page}
 %     \ContinuedFloat
     \caption{{\textbf{An example of a burst %with three peaks 
     for $n_{BK}=5$ %5 BK channels, 
     with 1:4 stoichiometry and $r=30$ nm}}. It corresponds to the burst of Fig.~\ref{fig:nBK5}f for $3.68 \le t \le 3.91$. Legends as in Fig.~\ref{fig:spike}. 
     Note how the system returns to the $m_{BK}=0$ plane (first subplot of panel (c)) to the right of the $V$-nullclines the first two times (first time, green points to the right of the solid light blue line for $Ca_c=0.4$; second time, black points to the right  of dashed blue line with $Ca_c=0.45\,\mu$M), leading to a new increase of $V$. 
     Only the third time, when $Ca_c$ has increased, moving the $V$ nullcline downwards, does the system return to the $m_{BK}=0$ plane to the left of the $V$ nullcline (brown points to the left of the dotted blue line with $Ca_c=0.49\,\mu$M), which causes a further decrease of $V$ terminating the burst. Note how the oval unstable limit cycle for $m_{BK}=0$ reduces by increasing $Ca_c$, approaching to the infinitesimal cycles with centres (-17 mV, 0.24) and (-18 mV, 0.23) for $Ca_c=0.45$ and $Ca_c=0.49\,\mu$M, respectively.}
\label{fig:burst_c}
%\end{center}
\end{figure*}

\begin{figure*}[htbp]
\centering
%\begin{center}
\subfigure[]{\includegraphics[width=0.35\textwidth]{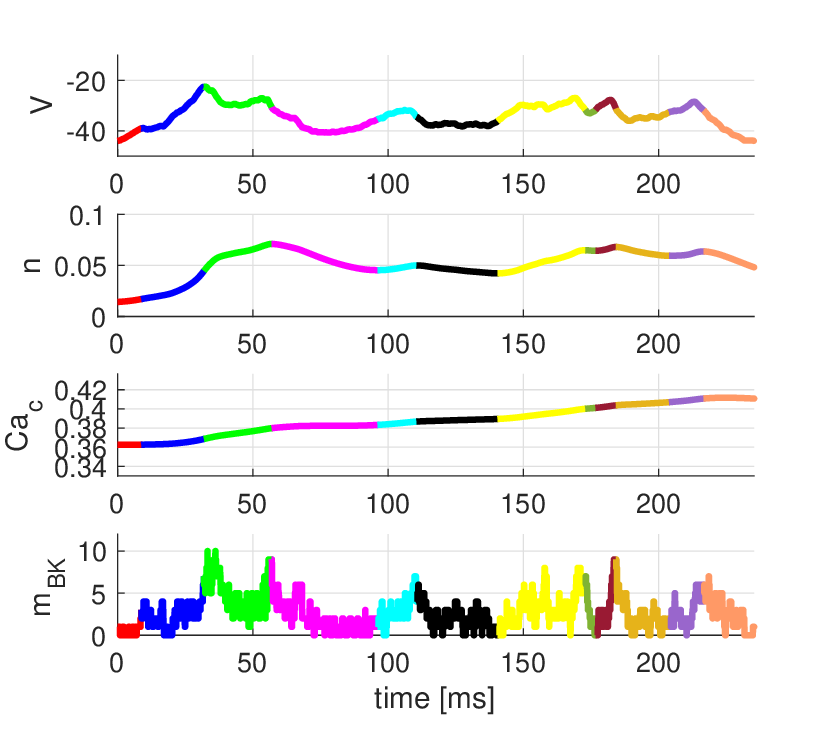}}
\subfigure[]{\includegraphics[width=0.35\textwidth]{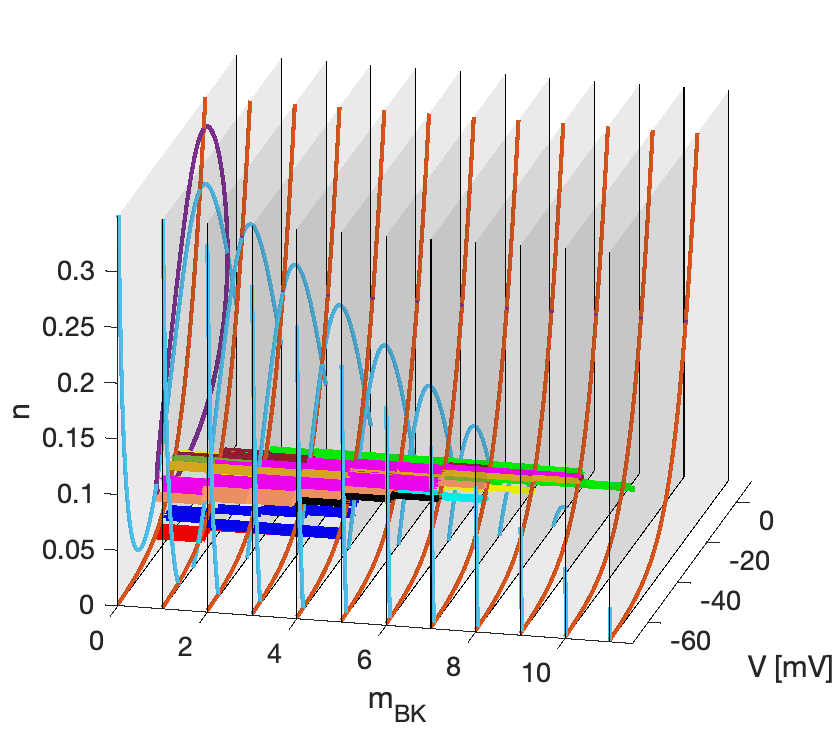}}
\subfigure[]{\includegraphics[width=0.7\textwidth]{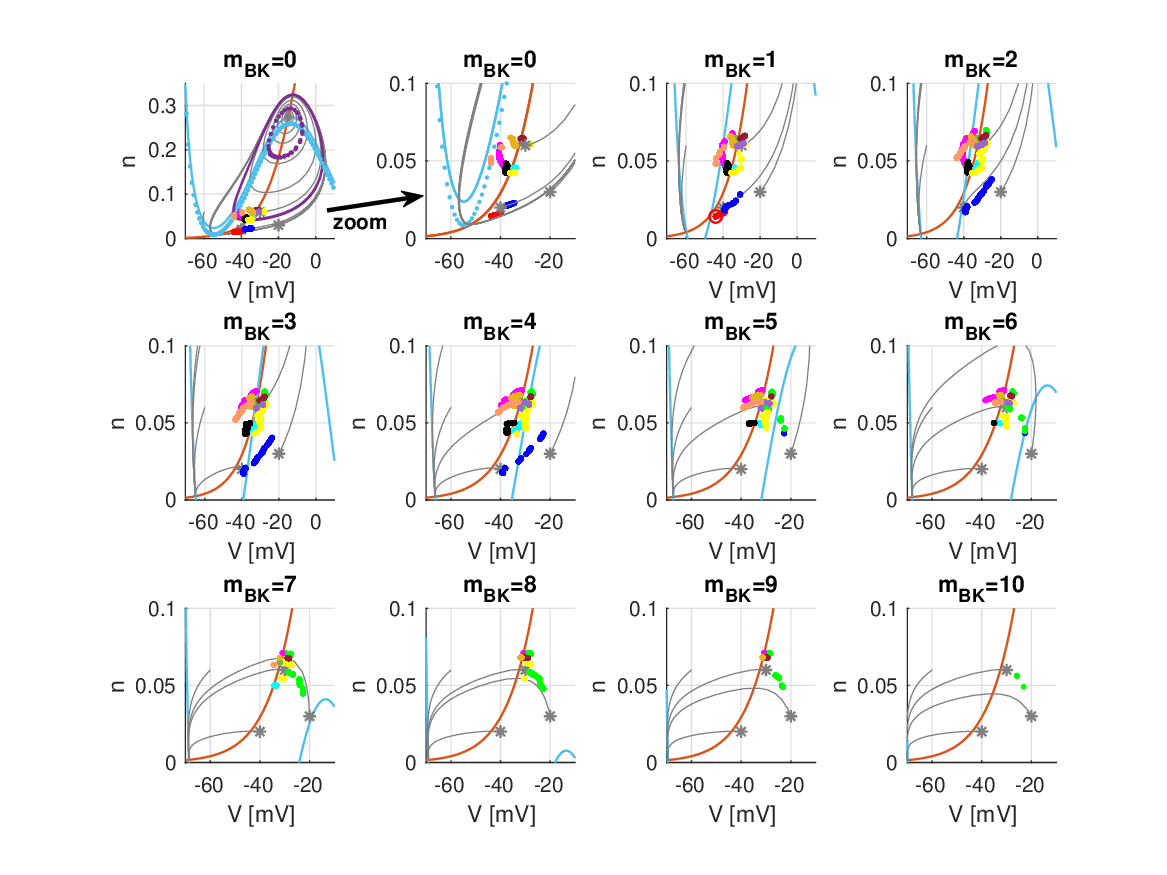}}
%\includegraphics[width=0.89\textwidth]{Figs/example_spike1_pp}
 % \caption[]{\textit{(see next page)}}
 % \end{figure}
 % \begin{figure} [t!]
 %     \captionsetup{labelformat=adja-page}
 %     \ContinuedFloat
     \caption{{\textbf{An example of a burst %with more peaks 
     for $n_{BK}=$ 15 %BK channels, 
     with 1:4 stoichiometry and $r=13$ nm}}. 
     It corresponds to the burst of Fig.~\ref{fig:nBK15}e for $0.93 \le t \le 1.16$. Legends as in Fig.~\ref{fig:spike}. 
     Note how the system returns to the $m_{BK}=0$ plane (first subplot of panel (c) and the relative zoom in reported in the second subplot) to the right of the $V$-nullclines more times and within the unstable limit cycle  
     (see magenta, cyan, black, yellow, green, brown and violet points to the right of the solid light blue line for $Ca_c \approx0.38$), leading to additional oscillations of $V$ from -40 to -25 mV. 
     Only, when $Ca_c$ has increased, moving  the $V$ nullcline downwards, does the system return to the $m_{BK}=0$ plane to the left of the $V$ nullcline (orange points to the left of the dotted blue line with $Ca_c=0.41\,\mu$M), which causes a further decrease of $V$ terminating the burst. Note how the oval unstable limit cycles for $m_{BK}=0$ reduce by increasing $Ca_c$: compare the greater violet oval for $Ca_c=0.38\,\mu$M with the smaller one for $Ca_c=0.41\,\mu$M. }
\label{fig:burst_d}
%\end{center}
\end{figure*}

\new{This observation implies for example that with only a single CaV located in the BK-CaV complexes, 
%(1:1 stoichiometry and $r=13$ nm), 
less BK channels will open as the cell depolarizes during the beginning of the action potential, compared to the case with 1:4 stoichiometry. % and $r=13$ nm (Fig. 
For example, in Fig.~\ref{fig:burst_b} (1:1 stoichiometry and $r=13$ nm), only 1-2 BK channels are open during most of the first increase in $V$ (red points), in contrast to the 2-4 BK channels being open during the upstroke in Fig.~\ref{fig:burst} (1:4 stoichiometry and $r=13$ nm). 
This implies that the system is below the $V$ nullcline for a longer time, leading to a first peak at $V\sim 0$ mV, compared to the peak at $V\sim -10$ mV for the case of 1:4 stoichiometry. 
Moreover, during the beginning of the downstroke (blue points in Fig.~\ref{fig:burst_b}c) the system is relatively close to the $V$ nullcline, meaning that $V$ does not decrease rapidly. 
Altogether, the long time that the systems spends at very depolarized $V$, allows the variable $n$ to increase more than in the previous case.  
The result is that when BK channels eventually close and the system reaches the $m_{BK}=0$ plane, it will often fall within the unstable limit cycle and begin to spiral counter-clockwise. Even if one or two BK channels open, the system is still in the region of the $(V,n)$ plane where the unstable limit cycle is lying in the $m_{BK}=0$ plane. Only when $Ca_c$ has increased sufficiently, so that the $V$ nullcline has moved sufficiently downwards, is the system able to escape and terminate the burst.
To obtain spiking, the system must fall outside and to the left of the unstable limit cycle when reaching the $m_{BK}=0$ plane, which is confirmed by our simulations (not shown).}

\new{To summarize, compared to the 1:4 stoichiometry scenario, the fact that the opening probability for BK channels is lower causes $n$ to increase so much that the system often falls on the inside of the unstable limit cycle when returning to the $m_{BK}=0$ plane. This mechanism explains why bursting is more frequent with fewer CaVs in the BK-CaV complexes for $r=13$ nm, compare panels (a) and (e) in Fig.~\ref{fig:nBK5}.
}

\new{A similar explanation underlies the increased propensity for bursting when CaVs are located 30 nm (rather than 13 nm) from the BK channel in the complex with 1:4 stoichiometry (panels (e) and (f) in Fig.~\ref{fig:nBK5}).
Again, BK channels tend to open less during the upstroke so that $n$ can increase more, which eventually makes $V$ decrease. As $V$ decreases, the BK channels tend to close and the system reaches the $m_{BK}=0$ plane at fairly high $n$ values (Fig.~\ref{fig:burst_c}) so that the system might fall inside or very close to the unstable limit cycle, but to the right of the $V$ nullcline, thus creating small-amplitude oscillations (Fig.~\ref{fig:burst_c}). 
This mechanism is similar to the one described producing bursting for $r=13$ nm and 1:4 stoichiometry, but it is more frequent for $r=30$ nm since $n$ typically will be higher at the action potential peaks, compare the red traces in panels (e) and (f) in Fig.~\ref{fig:nBK5}. }
%\fbox{Is all this true??}}

\new{With more BK channels in the cells, a higher number of BK channels will be open in spite of the same opening probability. 
For example, with $n_{BK}=15$ instead of $n{BK}=5$, three times as many BK channels will be open on average, in spite of all other configurations (membrane potential $V$, number of CaVs per BK-CaV complex, and CaV-to-BK distance $r$) being identical. 
Thus, during the upstroke, 
the system quickly reaches 
%the planes corresponding to 
$m_{BK}>5$, and the many open BK channels stop the increase in $V$ and lead to a first peak at $V< -20$ mV (Fig.~\ref{fig:burst_d}).
As a consequence of the relatively weak depolarization of the membrane potential, few Kv channels become activated, i.e., $n$ remains low ($n<0.08$ in Fig.~\ref{fig:burst_d}).
Geometrically, this means that the system remains in the lower part of the $(V,n)$ planes. Consequently, when the BK channels close during the downstroke and the system reaches the $m_{BK}=0$ plane, it will be to the right of the $V$ nullcline (Fig.~\ref{fig:burst_d}c), and the system will produce additional oscillations going from $V\approx -40$ mV to $V\approx -25$ mV, until $Ca_c$ increases sufficiently to move the $V$ nullcline downward so that the system falls in the region above the $V$ nullcline. 
In contrast to the previous cases, this occurs for much lower $n$ values, and hence the system does not follow the trajectories that reach a minimum for $V$ at $-60$ mV.}

\section{Discussion}
\label{discussion}
Understanding how complex dynamics arises in nonlinear, hybrid deter\-ministic-stochastic models is important for providing insight into the role of underlying mechanisms and for controlling the corresponding systems.
Geometrical methods have proven highly useful for deterministic systems, also in biology \cite{keener09,fall02,izhikevich07}.
Whereas deterministic models driven by continuous stochastic processes, e.g., Wiener processes, have been quite extensively studied \cite{devries00,pedersen05b,berglund06}, this is not the case for discrete hybrid systems. 
However, some results do exist for systems driven by a Markov chain independent of the other (deterministic) variables, so called blinking systems \cite{hasler13,barabash20}.
 
The model of electrical activity in a pituitary cell with stochastic ion channels that we studied here does not fall within either of the examples mentioned above, but is a truly random dynamical system \cite{arnold98}, since the transition rates of the stochastic variable ($m_{BK}$) depend on the deterministic variables ($V$). 
Recent studies have studied similar hybrid models of stochastic electrical activity by analyzing the corresponding ``average" deterministic model with geometric tools, and then interpreting ion channel noise as random perturbations (``pushes") of the deterministic system \cite{richards20,fazli21}. 

In contrast, here we have shown how one can work directly with the hybrid system. This is achieved by analyzing the hybrid (fast subsystem) phase space as the union of discrete planes, each one corresponding to a certain value of the discrete stochastic variable $m_{BK}$ indicating the number of open BK channels.
In the terminology of random dynamical systems, each of these planes is a \emph{fiber}, and the random part of the model corresponds to jumps between these fibers.
We showed that the locations of geometric structures, in particular nullclines and an unstable limit cycle, govern the behavior for fixed $m_{BK}$, 
and that the overall dynamics, e.g., whether the model produces a spike or a burst, is determined by the location at which the system jumps from one plane to another, in particular, the point at which the system reaches the $m_{BK}=0$ plane plays an important role.

%\fbox{slow-fast model...}
To reach this description, we took advantage of the slow-fast structure of the model. Since the \Ca\ variable $Ca_c$ operates on a slower timescale than the other variables, it can  be treated as a (slowly varying) parameter in the fast subsystem. 
For our model, this assumption has the big advantage that the fibers corresponding to fixed $m_{BK}$ values become two-dimen\-sion\-al, which helps geometrical reasoning.
The slow dynamics of $Ca_c$ was taken into account by considering how the relevant geometrical structures, in particular the $V$-nullcline, move as $Ca_c$ changes.

The strength of our approach is maybe best exemplified by its ability to explain counter-intuitive numerical results.
We noted that spiking is more frequently seen both when BK channels are located in complexes with four CaV channels at a distance of $r=13$ nm (Fig.~\ref{fig:nBK5}e), which would lead to a high BK open probability, and when BK channels are associated with just a single CaV at a large distance ($r=30$ nm, Fig.~\ref{fig:nBK5}c), corresponding to a low BK open probability. 
Thus, there seems to be a window of intermediate BK opening probabilities where bursting is favored.
We explained this by noticing that if the BK channels open to readily during the upstroke, the gating variable $n$ does not increase very much since the $V$ variable begins to decrease early. When returning to the $m_{BK}=0$ plane the system is often below that unstable limit cycle, but at $n$ large enough to be above the $V$ nullclines, which ends the action potential.
If the BK channels open more slowly, $V$ and consequently $n$ increase more, so that the system returns to the $m_{BK}=0$ plane near or even inside the unstable limit cycle, thus creating a second action potential, i.e., a burst, before hyperpolarizing.
In contrast, if the BK channel do not open sufficiently, so that $m_{BK}\leq 1$ except on rare occations (see Fig.~\ref{fig:nBK5}c), the system basically follows the stable limit cycle, corresponding to spiking, in the $m_{BK}=0$ plane, and the large orbit in the $m_{BK}=1$ plane (see gray curves in, e.g., Fig.~\ref{fig:spike}c).
Biologically, this suggests that BK-CaV complexes could be appropriately tuned %in order 
to increase the burst frequency.

Our analysis can also explain geometrically the observations by Richards et al.~\cite{richards20} that opening a BK channel just before the action potential peak increases the probability of observing a burst, whereas opening a BK channel during the downstroke of the action potential reduces the chance of producing a burst.
If a BK channel opens at high $V$, the $V$ nullcline moves down and $V$ will stop increasing and begin decreasing, which leads to less activation of Kv channels (smaller $n$). The results is that  the system returns to the $m_{BK}=0$ plane at lower $n$ values, where it is more likely to fall below and on the right of the middle branch of the $V$ nullcline.
In contrast, if a BK channel opens during the downstroke when both $V$ and $n$ are decreasing, the downward shift of the $V$ nullcline will accelerate the decrease of $V$, so that when the system eventually reaches the $m_{BK}=0$ plane, it will more likely hit above and to the left of the middle part of the $V$ nullcline.

In summary, we have presented a -- to the best of our knowledge -- novel method that combines ideas from standard geometrical analysis of smooth dynamical systems with a picture taken from random dynamical systems, where stochastic events correspond to random jumps between fibers. This approach, which successfully allowed us to explain complex behavior in a hybrid model of electrical activity influenced by stochastic ion channel dynamics in a pituitary cell, should be useful for similar models in cell biology as well as for other applications of nonlinear, hybrid models.

%\begin{acknowledgements}
%If you'd like to thank anyone, place your comments here
%and remove the percent signs.
%\end{acknowledgements}

% Authors must disclose all relationships or interests that 
% could have direct or potential influence or impart bias on 
% the work: 
%
\noindent \textbf{Statements and Declarations}\\
\noindent \textit{Funding:}
The authors declare that no funds, grants, or other support were received during the preparation of this manuscript.

\noindent\textit{Competing interests:}
The authors declare that they have no conflicts of interest.

\noindent \textbf{Data availability}\label{sec:data}\\
MATLAB/SIMULINK code to reproduce the results in the manuscript is available at 
\url{https://researchdata.cab.unipd.it/id/eprint/1029}
%s online supplementary material.

% BibTeX users please use one of
%\bibliographystyle{spbasic}      % basic style, author-year citations
% \bibliographystyle{spmpsci}      % mathematics and physical sciences
% %\bibliographystyle{spphys}       % APS-like style for physics
% %\bibliography{}   % name your BibTeX data base
% \bibliography{refs}

% Non-BibTeX users please use
% \begin{thebibliography}{}
% %
% % and use \bibitem to create references. Consult the Instructions
% % for authors for reference list style.
% %
% \bibitem{RefJ}
% % Format for Journal Reference
% Author, Article title, Journal, Volume, page numbers (year)
% % Format for books
% \bibitem{RefB}
% Author, Book title, page numbers. Publisher, place (year)
% % etc
% \end{thebibliography}

\end{document}